\documentclass[pdflatex,sn-mathphys]{article}

\usepackage[english]{babel}
\usepackage{subcaption}
\usepackage[a4paper]{geometry}
\usepackage{authblk}

\usepackage{adjustbox}
\usepackage{booktabs}
\usepackage{amsmath}
\usepackage{graphicx}
\usepackage[colorlinks=true, allcolors=blue]{hyperref}

\title{The limits of human mobility traces to predict the spread of COVID-19}

\author[1,2]{Federico Delussu}
\author[1,3 $\dagger$]{Michele Tizzoni}
\author[1,4$\dagger$]{Laetitia Gauvin}
\affil[1]{ISI Foundation, via Chisola 5, 10126, Turin, Italy}
\affil[2]{Department of Applied Mathematics and Computer Science, DTU, Copenhagen, Denmark}
\affil[3]{Department of Sociology and Social Research, University of Trento, Trento, Italy}
\affil[4]{Institute for Research on Sustainable DevelopmentIRD, UMR 215 Prodig, 5 cours des Humanités, F-93 322 Aubervilliers Cedex,
France}
\affil[$\dagger$]{\small these authors contributed equally to this work}

\date{}

\begin{document}

\maketitle

\begin{abstract}

Mobile phone data have been widely used to model the spread of COVID-19, however, quantifying and comparing their predictive value across different settings is challenging. Their quality is affected by various factors and their relationship with epidemiological indicators varies over time.
Here we adopt a model-free approach based on transfer entropy to quantify the relationship between mobile phone-derived mobility metrics and COVID-19 cases and deaths in more than 200 European subnational regions. 
We found that past knowledge of mobility does not provide statistically significant information on COVID-19 cases or deaths in most of the regions. In the remaining ones, measures of contact rates were often more informative than movements in predicting the spread of the disease, while the most predictive metrics between mid-range and short-range movements depended on the region considered.  
We finally identify geographic and demographic factors, such as users' coverage and commuting patterns, that can help determine the best metric for predicting disease incidence in a particular location. 
Our approach provides epidemiologists and public health officials with a general framework to evaluate the usefulness of human mobility data in responding to epidemics.

\end{abstract}

\section{Introduction}
The relationship between human movements and the spatial spread of infectious diseases has been recognized for a long time \cite{longini1988mathematical,findlater2018human,balcan2010modeling}. Human movement has been shown to play a key role in the dynamics of several pathogens, through two basic mechanisms: traveling infectious individuals may introduce a pathogen in a susceptible population, and, at the same time, human movement increase the contact rate between individuals, creating new opportunities for infection.
In the past 15 years, the increasing availability of mobility data derived from mobile phones has fueled a large body of work aimed at identifying opportunities to use them for infectious disease modeling and surveillance \cite{wesolowski2016connecting,wesolowski2012quantifying,mari2012modelling,buckee2013mobile,charu2017human,tizzoni2014use,peak2018population}.

More recently, during the COVID-19 pandemic, mobile phone-derived data have been extensively harnessed to monitor the effect of non-pharmaceutical interventions (NPIs) across countries, understand the early dynamics of COVID-19 diffusion, and forecast its spread at different spatial scales, from countries to cities ~\cite{zhang2022human,oliver2020mobile,buckee2020aggregated, gatto2020spread, cramer2022evaluation,chang2021mobility,lucchini2021living}.
By measuring human movements and combining them with phylogeography methods~\cite{lemey2014unifying, lemey2020accommodating}, several studies shed light on the cryptic spread of new variants, their persistence over time and resurgence after the relaxation of NPIs \cite{kraemer2021spatiotemporal, davis2021cryptic,lemey2021untangling}.

Human mobility has been shown to strongly correlate with the spread of COVID-19 during the early phase of the outbreak in China and in many other countries \cite{chinazzi2020effect,peixoto2020modeling, kraemer2020mapping,jia2020population,persson2021monitoring,iacus2020human}.
However, once COVID-19 established a foothold in a population, the relative importance of mobile phone-derived data to predict the epidemic dynamics on a local scale has been generally less understood and several studies have shown conflicting evidence about the use of mobility traces to model the spread of COVID-19 at later stages of the outbreak.
For instance, it has been shown that the explanatory power of mobility metrics in relation to the case growth rate in the U.S., significantly declined in spring 2020, especially in rural areas \cite{kishore2021evaluating, jewell2021s, badr2021limitations}. 
Similar trends have been observed in Europe \cite{nouvellet2021reduction}.
In parallel, mobile phone-derived data have been proven beneficial to model COVID-19 dynamics in largely populated urban areas of Western countries \cite{aleta2020modelling, aleta2022quantifying}, but less so in countries of the Global South \cite{ramiadantsoa2022existing}.

Several reasons have been proposed to explain the varying relationship between mobility metrics and epidemic indicators \cite{kishore2021evaluating}.
Mobility metrics are generally derived from raw mobile positioning data through complex and customized processing pipelines that can significantly vary across data providers \cite{kishore2021mobility}. 
How raw data are processed, and the specific definitions of mobility metrics can significantly impact their interpretation with respect to epidemic variables \cite{levin2021insights}.
Moreover, the relationship between mobility and epidemic patterns often relies on modeling assumptions, typically considering linear dependencies, that may not capture the complex interplay of these quantities \cite{nouvellet2021reduction, jewell2021s}.
Finally, mobile phone-derived metrics are generated from a sample of users who is generally not representative of the whole population.
It is therefore of paramount importance to define standardized approaches that can quantify the added value of mobility metrics for epidemiological analysis, and make different metrics, across settings, directly comparable.

Here, we extensively quantify the relationship between cell phone-derived mobility metrics and COVID-19 epidemiological indicators through a model-free approach, based on an information-theoretic measure, transfer entropy \cite{schreiber2000measuring}, adapted for small sample sizes.
Leveraging granular data provided by Meta that capture both users' movements and colocation at a fine spatial scale \cite{iyer2020large}, we measure the information flow between mobility metrics and time series of COVID-19 incidence and deaths in four European countries, at a subnational scale, over a one year period.
We find that the relative information added by the past knowledge of mobility metrics to the knowledge of the current state of COVID-19 time series is 
often not statistically significant. 

In statistically significant cases instead, we show that the relative information added by past knowledge of COVID-19 cases to the knowledge of current deaths is twice the information flow between past knowledge of mobility metrics and current deaths. 
We also show that the information flow of a given mobility metric to predict future COVID-19 incidence or deaths can be significant in one country but not in another, even if derived from the same original data source.

Being a general framework, our approach provides a quantitative measure of the relative added explanation brought by mobile phone data to the prediction of epidemiological time series that does not depend on the choice of a specific forecasting model. It thus helps to better identify the most appropriate mobility metrics to use among those available.
Our results can thus guide epidemiologists and public health practitioners in the evaluation of mobile phone-derived mobility metrics when they are interpreted as a precursor of epidemic activity.

\section{Results}

Here, we first describe and then apply our framework to measure the information flow between human mobility traces and the time evolution of COVID-19 in four European countries. 

\subsection{A transfer entropy approach to link mobility behavior and COVID-19 epidemiology} 

\begin{figure}[tb!]
\centering
\includegraphics[width=0.9\textwidth]{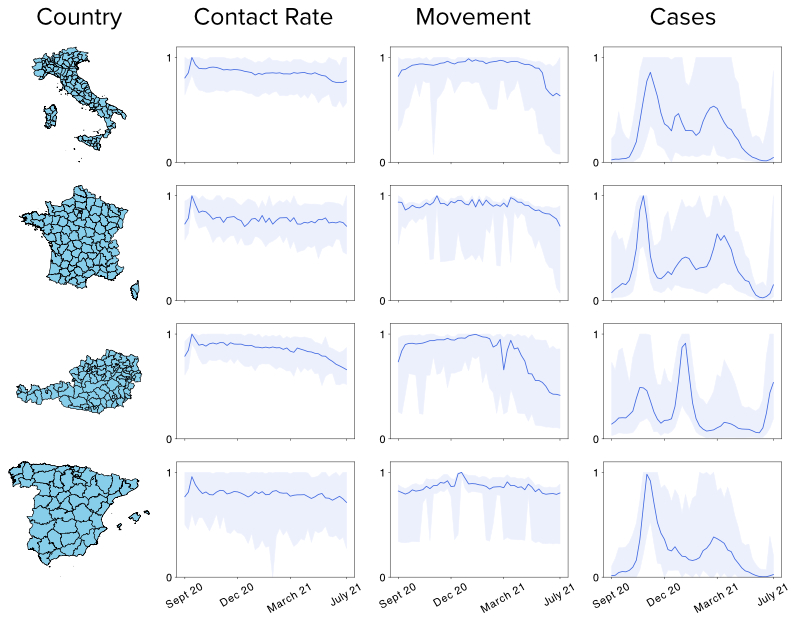}
\caption{\label{fig:datasets} \textbf{Summary of behavioral and epidemiological indicators.} In each country under study (from top to bottom: Italy, France, Austria and Spain), we consider three different types of indicators: contact rates, movements (here for the sake of simplicity we only show the short-range movements), and COVID-19 cases. 
In each plot, the blue shaded area highlights the within-country variability, corresponding to time series in every administrative subdivision. The blue solid line represents the average value. All curves are normalized between 0 and 1, corresponding to their maximum value. 
}
\end{figure}

With the aim of quantifying the information flow from mobility-derived data to COVID-19 data, we first gathered a set of mobility and epidemiological indicators.
Fig.~\ref{fig:datasets} provides an overview of the datasets used in the study. In Materials and Methods, we provide a full description of all data sources and the data processing steps.
We considered four European countries -- Austria, France, Italy, and Spain -- and their administrative subdivisions at NUTS3 level \cite{nuts2020europe} which is the lowest, i.e. the most granular, level of the standard hierarchy of administrative regions in Europe (Fig.~\ref{fig:datasets}, leftmost column). 

In all administrative regions, we collected indicators of the COVID-19 epidemic dynamics, namely, the weekly and daily numbers of new COVID-19 cases and deaths over the period, from September 2020 until July 2021.
During this period, the dynamics of COVID-19, exemplified by the incidence of new cases (Fig.~\ref{fig:datasets}, rightmost column), displayed subsequent waves, as a result of the complex interaction between the spread of new variants, the adoption of non-pharmaceutical interventions, the introduction of vaccines. 

In each country, we also collected weekly and daily time series describing movements and colocation patterns made available by Meta \cite{Facebook_data}.
We computed contact rates from colocation maps (see Material and Methods and the SI for details), which measure the probability that two users from two locations are found in the same location at the same time \cite{iyer2020large}. 
Colocation maps were generated by Meta on a weekly basis, only.
To study human movement patterns, we considered movement range maps provided by Meta, which report the number of users who moved between any two 16-level Bing tiles with an 8 hour frequency \cite{HDX}. 
To make colocation and movement patterns comparable in terms of scale, we focused on short-range movements, i.e. movements that occurred within the same tile, and we separately considered the mid-range movements, i.e. movements that occur between two different tiles in the same province.

We then processed the three datasets, starting from their raw form, to aggregate them at the NUTS3 resolution and create the time series: $M^s(t)$ for the short-range movements, $M(t)$ for the mid-range movements and $CR(t)$ for the contact rates. 
These time series were then used as source variables in the information-theoretic analysis. 
In the remainder of the paper, we will generally refer to $CR(t)$, $M^s(t)$, and $M(t)$ as mobility time series as they are all derived from human mobility data. 
We will also generally refer to the NUTS3 units as provinces, although their nomenclature varies across countries.

\begin{figure}[tb!]
\centering
\includegraphics[width=0.9\textwidth]{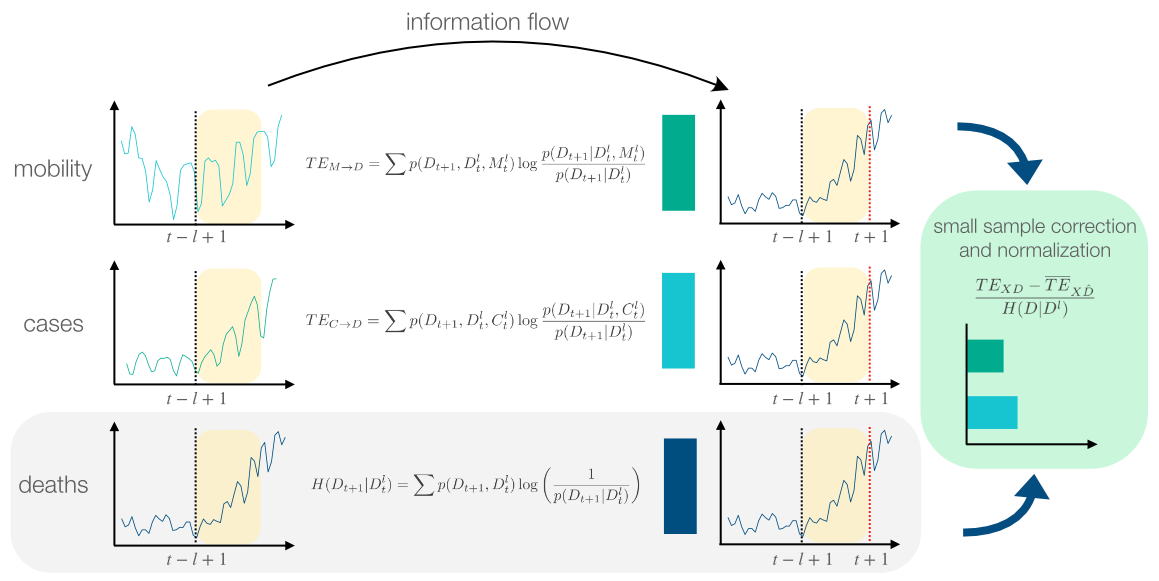}
\caption{\label{fig:tentropy}\textbf{Illustration of study design.} We computed the transfer entropy $TE_{X \rightarrow Y}$ to measure the information flow between source $X$ (on the left) and target time series $Y$ (right), for a given time lag $l$. In the figure example, as target time series we consider the number of COVID-19 deaths, $D(t)$. As source time series, we consider either mobility indicators, $M^s(t)$, $M(t)$, $CR(t)$, or COVID-19 cases $C(t)$. Transfer entropy quantifies the amount of information that is added by past knowledge of mobility or cases (green and cyan bars, respectively) to current knowledge of deaths, with respect to the knowledge of past deaths only (blue bar). 
After correcting the TE for small sample sizes, and normalizing by the reference value represented by the blue bar, we finally compare the Normalized Effective Transfer Entropy of mobility and cases (rightmost box). 
}
\end{figure}

Fig.~\ref{fig:tentropy} illustrates our study design based on the transfer entropy \cite{schreiber2000measuring}. 
Transfer entropy is a metric that measures the directed statistical dependence between a source and a target time series and it has been applied to a wide range of research domains \cite{bossomaier2016transfer}. 
Here, our approach consists, first, in computing the transfer entropy between mobility time series, $M^s(t)$, $M(t)$ and $CR(t)$, and epidemiological time series such as the reported number of COVID-19 attributed deaths $D(t)$ and cases $C(t)$, in each administrative unit, and for different temporal lags $l$, using the definition of Shannon entropy, as described by the equations in Fig.~\ref{fig:tentropy}.
Intuitively, the transfer entropy between mobility and deaths, $TE_{M^s \rightarrow D}$ (resp. $TE_{M \rightarrow D}$), can be interpreted as the degree of uncertainty of the reported deaths, $D$, at time $t$ that is solved jointly by the time series of deaths and mobility trends $M^s$ (resp. $M$) and exceeds the current degree of uncertainty of $D$, which can be solved by $D$'s own past.

It is known that transfer entropy estimates suffer in case of small sample sizes and non-stationarity of the source and target time series~\cite{marschinski2002analysing}. 
Moreover, due to the non-parametric nature of the transfer entropy, values computed between different source-target time series are not directly comparable.
To address these issues, we first adopted the definition of effective transfer entropy (ETE)~\cite{marschinski2002analysing}. 
ETE is obtained by subtracting from the original definition of TE a reference TE value using a shuffled version of the target time series (see Methods for details), thus removing spurious contributions to TE due to fluctuations observed in small sample sizes. 
Also, to address biases due to small sample sizes, we applied a Kernel Density Estimation, before the time series discretization that is necessary to compute the transfer entropy. 
Second, we normalized the effective transfer entropy by the Shannon entropy of the target variable, defining a normalized effective transfer entropy (NETE) \cite{zeng2022spacecraft}. 
We obtain a metric that is always positive when it is statistically significant and whose zero value indicates the absence of information transfer between time series. 
In the remainder of the article, we thus refer to the NETE between source $X$ and target $Y$ as our main quantity of interest, using the symbol $N_{X \rightarrow Y}$ to denote it. 

To better understand the cause-effect relationship between mobility and COVID-19 deaths, which are encoded in the value of $N_{M\rightarrow D}$ ,$N_{M^s \rightarrow D}$ and $N_{CR \rightarrow D}$, we compared them against the transfer entropy $N_{C \rightarrow D}$, where $C$ is the time series of new COVID-19 cases.
As the causal relationship between the number of cases and deaths is established by definition, we used the transfer entropy $N_{C \rightarrow D}$ as a benchmark to evaluate the added value of mobility indicators to predict COVID-19 deaths.
As an example, similar values of $N_{M^s \rightarrow D}$ and $N_{C \rightarrow D}$ would suggest knowledge of past COVID-19 incidence encodes a similar amount of information as knowledge of past mobility when it comes to predicting future deaths.

\subsection{The information flow between COVID-19 incidence and deaths}

As previously mentioned, to gauge our transfer entropy analysis framework, we first looked at the causal relationship between the incidence of COVID-19 cases and reported death counts.
It is clearly expected that a major source of information that provides knowledge on future deaths is encoded in the time series of past case counts. 
We used the NETE to quantify such information flow. 
\begin{figure}[htb]
\centering
\includegraphics[width=0.9\textwidth]{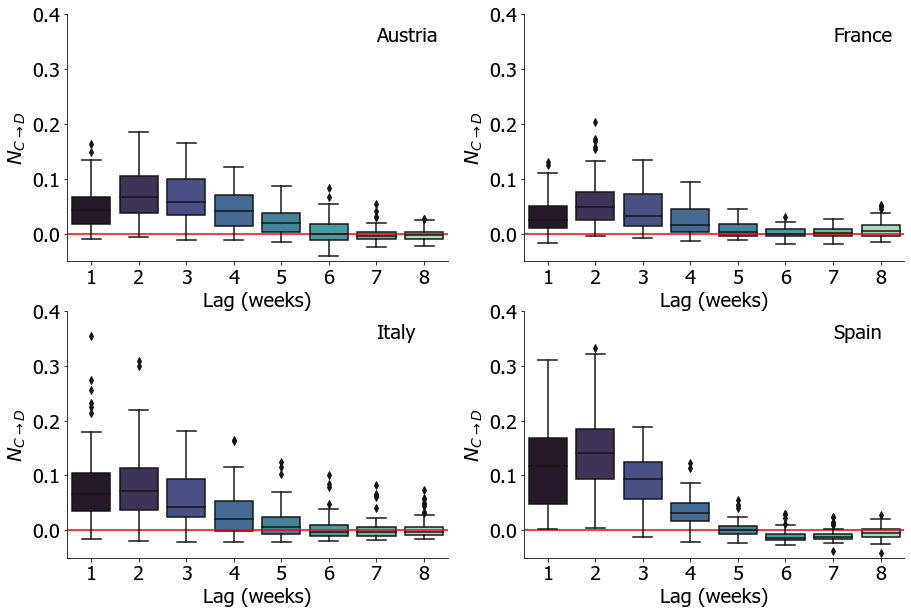}
\caption{\label{fig:TE_cases} 
\textbf{Information flow between COVID-19 incidence and deaths.}
Normalized Effective Transfer Entropy (NETE) between COVID-19 weekly reported cases and deaths in the NUTS3 administrative subdivisions  (provinces) of Austria, France, Italy and Spain. NETE is computed for lags ranging from 1 to 8 weeks, on the x-axis. Boxplots are computed on the distribution of NETE values of all the administrative subdivisions in each country. The horizontal red line marks the value $N_{C \rightarrow D} = 0$.
}
\end{figure}

Fig.~\ref{fig:TE_cases} shows the NETE between the weekly time series of COVID-19 cases and deaths in the four countries under study.
In all countries, median values of $N_{C \rightarrow D}$ increase from lags equal to 1 week up to a maximum of around 2-3 weeks, and then decline rapidly beyond the 3 weeks time lag. 
This is in line with early estimates of the median time delay between case reporting and fatality, which was estimated to range between 7 and 20 days in different countries \cite{wilson2020case, fritz2022wave}. 
At lag equal to 2 weeks, the mean relative explanation added by time series of cases with respect to deaths -- that is how much of $D(t)$ can be explained only by the past knowledge $C(t-l)$ -- is $14\%$ (SD=8) in Spain, $8\%$ (SD=6) in Italy, $7\%$ (SD=5) in Austria, and $6\%$ (SD=5) in France.
Boxplots computed on the distribution of administrative units in each country show a substantial heterogeneity of NETE across regions for lags shorter than 4 weeks. 
This may be partially explained by spatial heterogeneities in case and death reporting, and in testing strategies. 
Also, $N_{C \rightarrow D}$ values appear to be higher in Spain, with respect to the other countries.
A transfer entropy analysis of daily time series of COVID-19 cases and deaths displays consistent results (see Fig. S1), with NETE values that fall within the same range measured on a weekly time scale.

These results suggest NETE estimates are robust with respect to the time scale at which source and target time series are compared. Moreover, it provides a reference value for NETE, in terms of orders of magnitude, when the existence of a causal relationship between time series is known.

\subsection{The information flow between mobility traces and COVID-19 dynamics}

\begin{figure}[htb]
\centering
\includegraphics[width=0.8\textwidth]{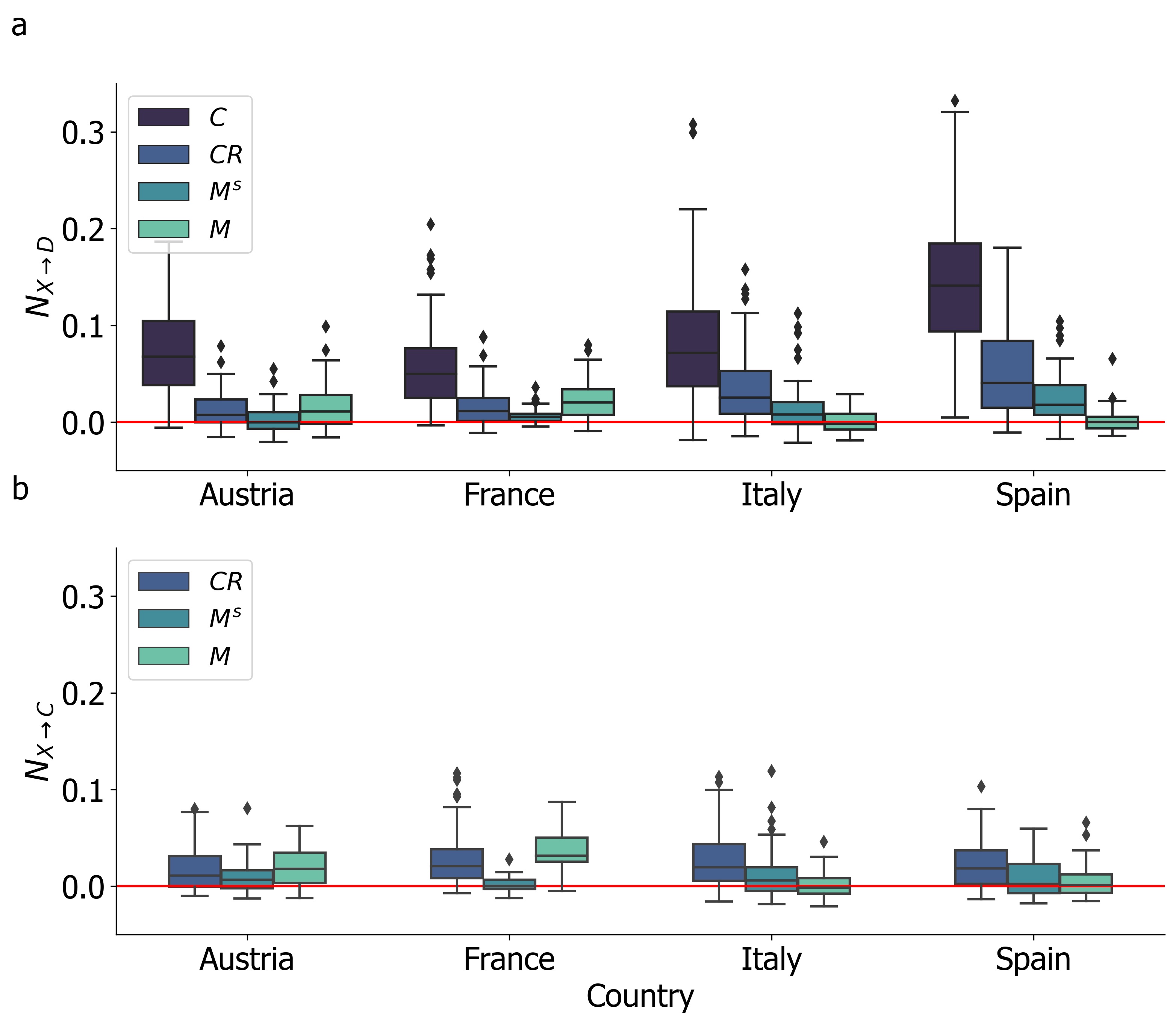}
\caption{\label{fig:TE_mobility} \textbf{Information flow from mobility data to COVID-19 incidence and deaths.} Comparison between the normalized effective transfer entropy computed from source time series $X$ and target time series of reported COVID-19 deaths $D$ (a) and cases $C$ (b).
Source time series are COVID-19 cases (only for deaths), contact rates, short range and mid-range movement. 
Boxplots are computed from the distribution of NETE values for a given time delay, $l$. In panel a: $l$= 2 weeks for cases, 7 weeks for contact rates and movement.
In panel b: $l$= 6 weeks for short range and mid-range movement.
The horizontal red line marks the value $N_{X \rightarrow D} = 0$.
}
\end{figure}

Having defined a benchmark of information transfer using $N_{C \rightarrow D}$, we measured the information flow between behavioral time series of mobility indicators and COVID-19 cases and deaths. 
Fig.~\ref{fig:TE_mobility} summarizes the main results of our analysis.
Values of $N_{X \rightarrow D}$, with $X$ being either short range movements, mid-range movements or contact rates, were substantially smaller than $N_{C \rightarrow D}$ in all countries, for any given time lag $l$.
In particular, Fig.~\ref{fig:TE_mobility}a allows to compare the distributions of $N_{C \rightarrow D}$, $N_{CR \rightarrow D}$, $N_{M^s \rightarrow D}$, and $N_{M \rightarrow D}$, at the time lag $l$ that maximized the median NETE for weekly time series, for all indicators.
We found the largest median values of the normalized transfer entropy at $l=7$ weeks for both contact rates and movements (short-range and mid-range). 
The upper quartile of the NETE distributions derived from the mobility traces generally fell below $5\%$, in all countries, while the lower quartile of $N_{C \rightarrow D}$ was always above $5\%$.
Also, the distributions of normalized transfer entropy computed from movements were much narrower and often including the value $N=0$ within their interquartile range.
Values of $N_{M \rightarrow C}$, shown in Fig.~\ref{fig:TE_mobility}b, display a pattern similar to the normalized transfer entropy from the mobility time series to the death time series, with generally low values of NETE in all countries. 
Compared to movement time series, contact rates led generally to relatively higher values of NETE with both targets, cases and deaths, as shown in Fig. \ref{fig:TE_mobility}.
Our result confirms the additional value of measuring contact rates from mobile phone data, with respect to other movement metrics \cite{crawford2022impact}. 
Besides, it shows that short-range mobility within a province had often a limited predictive power to capture time trends of COVID-19 spread.

\begin{table}[tb]
\centering
\begin{tabular}{c|lllllll}
\toprule
{} & \multicolumn{3}{l}{$\rightarrow C(t) (\%)$} & \multicolumn{4}{l}{$\rightarrow D(t) (\%)$} \\
$l$ (weeks) &    $CR(t)$&    $M(t)$ &    $M^{s}(t)$  &     $CR(t)$ &    $M(t)$  & $M^{s}(t)$  & $C(t)$ \\
\midrule
2   &   9 &  19& 3 &   10 & 7&  7 &  79 \\
3   &  20 &   23&5 &   21 & 8& 13 &  69 \\
4   &  27 & 22&  9 &   29 &  9&16 &  46 \\
5   &  33 & 23& 10 &   36 & 8&  17 &  18 \\
6   &  35 & 27& 10 &   38 & 14& 17 &   7 \\
7   &  29 &25&  11 &   40 & 12& 14 &   4 \\
8   &  27 & 20& 11 &   38 & 15& 12 &   8 \\
\bottomrule
\end{tabular}
\caption{\textbf{Percentage of statistically significant NETE values across provinces in all the countries studied}. This table shows the percentage of provinces, in all countries, in which the NETE is statistically significant ($p<0.01$) for lags ($l$) from $2$ to $8$ weeks. 
\label{NETE_number}}  
\end{table}

To obtain a more detailed picture of the predictive power of different mobility metrics in terms of NETE, we computed the percentage of provinces for which mobility time series provided significant relative information added, with respect to the past knowledge of epidemiological indicators only (see Tab. \ref{NETE_number}).
On the one hand, our framework effectively captured the existing causal relationship between the time evolution of cases counts and the number of deaths, as the NETE between these indicators was statistically significant ($p<0.01$) in about $80\%$ of the provinces, at 2 weeks lag.
On the other hand, we observed a statistically significant information transfer from mobility time series to epidemiological ones in a much smaller fraction of provinces.
Short-range movements NETE was significant in less than $20\%$ of provinces when considered as a predictor of both cases and deaths.
Mid-range movement time series and contact rates were significant in at most 27\% and 40\% of provinces. 
This means that in most provinces, mobility traces did not provide any additional information to predict future COVID-19 cases or deaths, at any lag between 2 and 8 weeks. 

Measures of contact rate extracted from colocation maps were more suitable than movement data to capture behavioral patterns relevant to predict COVID-19 spread.

By focusing only on those provinces where we could identify a significant information flow between mobility traces and COVID-19 indicators, we observe that the averaged relative explanation added by mobility data with respect to the epidemiological data ranges between $4-6\%$, which is about half of the averaged relative explanation added by past knowledge of cases to the prediction of future deaths (see Tab. \ref{NETE_perc} and Figs. S2-S9 in the SI). 

\begin{table}[tb]
      \centering
\begin{tabular}{c|lllllll}
\toprule
{} & \multicolumn{3}{l}{$\rightarrow C(t) (\%)$} & \multicolumn{4}{l}{ $\rightarrow D(t) (\%)$} \\
$l$ (weeks) &         $CR(t)$ &    $M(t)$ &       $M^{s}(t)$  &         $CR(t)$ &    $M(t)$ &       $M^{s}(t)$  &       $C(t)$ \\
\midrule
2 &  4 (1) & 4(1)&  4 (0) &  4 (1) & 5(2)& 4 (1) &  11 (6) \\
3 &  4 (2) &  4(2) &4 (1) &  5 (2) & 4(1)& 4 (1) &   9 (4) \\
4 &  5 (2) &  4(1)&4 (2) &  5 (2) & 4(1)& 5 (2) &   6 (3) \\
5 &  5 (2) &  4(1) &5 (2) &  6 (3) & 4(1)& 5 (2) &   5 (2) \\
6 &  6 (2) &  4(1) &5 (2) &  6 (3) & 4(2)& 5 (2) &   5 (2) \\
7 &  5 (2) &  5(1) &5 (2) &  6 (3) & 5(2)& 6 (3) &   5 (2) \\
8 &  5 (3) & 5(1)& 5 (2) &  6 (3) & 5(2) &6 (3) &   4 (1) \\
\bottomrule
\end{tabular}
     \caption{\textbf{NETE results across provinces in all the countries studied.} The table shows the average relative explanation added by source time series, with respect to past knowledge of the target only. Only provinces having a statistically significant NETE are considered. Numbers in parenthesis report the standard deviation computed over all provinces for which the NETE was statistically significant.}
      \label{NETE_perc}
\end{table}

As a sensitivity analysis, we also computed the NETE on a shorter time window, between September 2020 and January 2021, to exclude the confounding effect of the introduction of nationwide vaccination programs.
Since in those months all countries adopted mobility restrictions to mitigate the fall COVID-19 wave, we expect a stronger relationship between mobility and COVID-19 cases.
Indeed, during this time frame, the information flow between movement time series and COVID-19 cases was consistently higher than in the full study period (see Fig. S10). 
This result indicates that, provided with time series of adequate size, the NETE can effectively capture the time-varying relationship between human mobility time trends and COVID-19 dynamics.

\subsection{Identifying the determinants of mobility data predictive power for COVID-19
}

\begin{figure}[tb!]
\centering
\includegraphics[width=\textwidth]{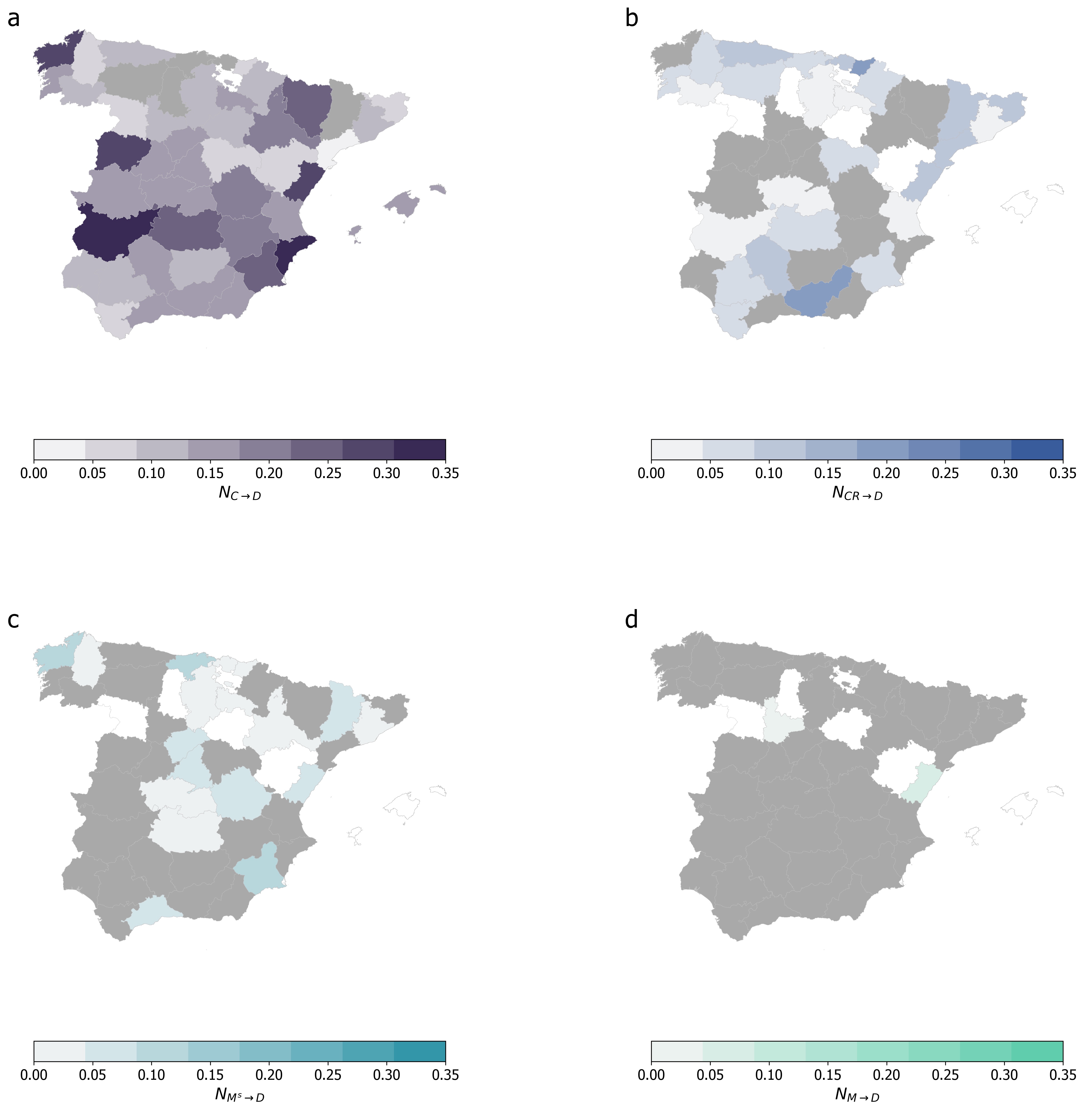}
\caption{\label{fig:map_Spain} \textbf{Spatial variations of normalized effective transfer entropy.} Maps of NETE values computed for different source time series and weekly COVID-19 deaths, in the provinces of Spain: (a) source is COVID-19 cases at lag $l$=2 weeks, (b) source is contact rate at lag $l$=7 weeks, (c) source is short-range movement at lag $l$=7 weeks. (d) source is mid-range movement at lag $l$=7 weeks.
Dark grey indicates provinces with non-significant values of NETE ($p>0.01$). Provinces in white are excluded from our sample. 
}
\end{figure}

Maps of Fig.~\ref{fig:map_Spain} highlight the spatial heterogeneity of $N_{X \rightarrow D}$ values observed within the same country, Spain, for a given time lag and different source time series (see Figs. S11 - S13 for the maps of Austria, France, and Italy). 
As previously mentioned, $N_{C \rightarrow D}$ displays higher and significant values in most of the country (Fig.~\ref{fig:map_Spain}a), with very few exceptions, while statistically significant values of $N_{M^s \rightarrow D}$ are found only in 16 provinces out of 42 (Fig.~\ref{fig:map_Spain}c).

To better understand the observed heterogeneity in NETE, and identify those features that can predict the likelihood to observe a statistically significant information transfer from mobility to COVID-19 death counts, we resorted to a classification model. 
Namely, we used a random forest classifier to predict when the value $N_{X \rightarrow D}$ is more likely to be statistically significant, using short-range movement and contact rate as source time series. We focused on these two metrics as they are quantities measured at the same spatial scale. Moreover, short-range movements represent on average $90\%$ or more of all movements within a province (see Table S1).
As input features to the model, we considered a set of attributes of the provinces in each country.
In particular, we investigated the effects of population size, province area in square kilometers, the density of Facebook users, the number of total cumulative deaths, the ratio between the number of commuters traveling from or to the province, and those who live and work there, as reported by the census (commuting flow), and the coverage consistency, that is the correlation  over time between the number of Facebook users sharing their location and the number of Facebook users taken into account to compute the colocation maps. 

The results summarized in Tab.~\ref{tab:classification} show that the model achieves a good overall performance in terms of precision and recall, as indicated by f1-scores generally higher than 0.6. 
In particular, of all provinces that are classified by the model as characterized by a statistically significant value of NETE, 90\% or more display a significant transfer of information, as shown by precision values.
On the other hand, the model's recall is close to 0.95 when it comes to identifying provinces characterized by a not statistically significant NETE, therefore the model correctly identifies 95\% of those provinces where there is no actual transfer of information between mobility and deaths. 

\begin{table}[tb]
  \begin{subtable}{.5\linewidth}
      \centering
      \begin{tabular}{lrr}
\toprule
{} &           p $\ge$ 0.01 &  p $<$0.01\\
\midrule
precision &    0.64&    0.90    \\
recall    &    0.95 &    0.47 \\
f1-score  &    0.77 &    0.62 \\
\bottomrule
\end{tabular}
      \caption{Movement}
    \end{subtable}%
    \begin{subtable}{.5\linewidth}
      \centering
      \begin{tabular}{lrr}
\toprule
{} &          p $\ge$ 0.01 &  p$<$0.01\\
\midrule
precision &    0.71 &    0.92\\
recall    &    0.95 &    0.61 \\
f1-score  &    0.81 &    0.74 \\
\bottomrule
\end{tabular}
 \caption{Contact rate}
    \end{subtable} 
     \caption{\textbf{Classification performance metrics.} Summary of model's classification performance to predict the statistical significance of NETE at the $p<0.01$ threshold when the input source is short-range movement (a) or contact rate (b) and target variable are COVID-19 deaths.}
     \label{tab:classification}
\end{table}

\begin{figure}[tb!]%
    \centering
    \subfloat[\centering Movement]{{\includegraphics[width=7.7cm]{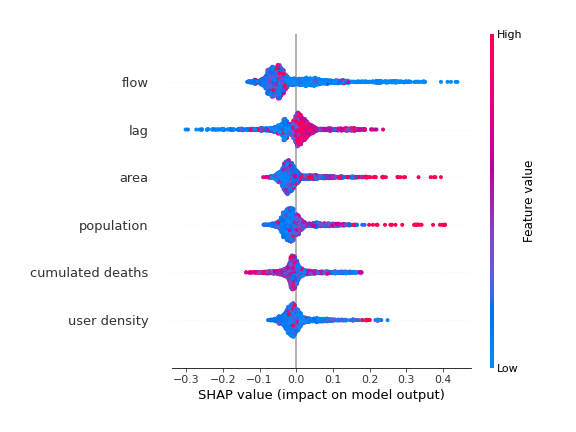} }}%
    \qquad
    \hspace{-1.8cm}
    \subfloat[\centering Contact rate]{{\includegraphics[width=7.7cm]{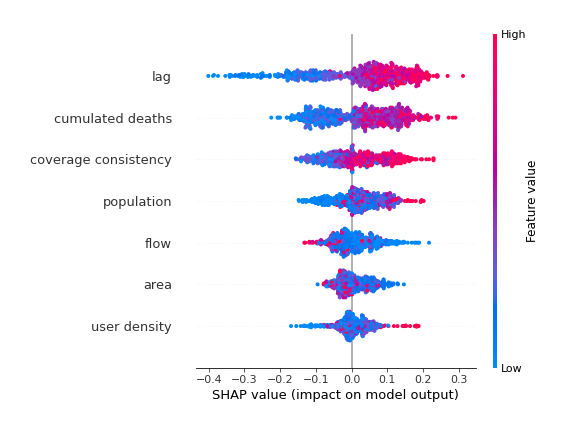} }}
    \caption{\textbf{SHAP plots of feature importance to predict the statistical significance of the NETE for all selected provinces.} Color represents the feature value (blue is low and red is high). Panel a describes the results for $N_{M^{s} \rightarrow D}$, panel b for $N_{CR \rightarrow D}$. The SHAP value, on the horizontal axis, indicates the feature importance on the model output, with larger values corresponding to higher relevance. Each dot represents a single observation. Features are ranked by importance.}%
    \label{fig:shap}
\end{figure}

To explore the importance of province features in our classification model, we examined the SHAP (SHapley Additive exPlanations) values associated with each, as shown in Fig.~\ref{fig:shap}. 
SHAP is a method based on a game theoretic approach to explaining the output of classification models~\cite{lundberg2017unified}.
As expected, the choice of the time lag to compute the NETE is crucial in determining the presence of a significant information transfer between mobility metrics and epidemiological indicators. Indeed, $lag$ is ranked as the most and second most important feature explaining the classification, for contact rate and short-range movement, respectively.
Commuting flow is the most important predictor of the statistical significance of NETE between short-range movements and deaths: when the number of commuters leaving or entering a province represents an important fraction with respect to those who remain within the province, the relationship between short-range mobility and COVID-19 dynamics gets weaker.  
However, the same feature has only a marginal impact on the NETE between contact rates and deaths, which suggests contact rate should be preferred over short-range movements to predict epidemic outcomes when a province is characterized by large population inflows/outflows.   
Province area and population size have also a significant impact on the information transfer between short-range movement and COVID-19 deaths. 
Indeed, a larger area and population size correspond to a higher likelihood of NETE significance for short-range movements.

This effect may partly explain why we observed NETE values that were statistically significant only in a few provinces of Austria, where spatial units were particularly small.
When looking at the information flow between contact rates and time series of deaths, the total cumulative deaths represent an important explanatory variable for the classification model. 
Besides the analysis presented in Fig.~\ref{fig:shap} suggests that the coverage consistency needs to be sufficiently high in order to get a statistically significant transfer entropy from contact rate to deaths. 
In France, where in most provinces the coverage consistency is low and the commuting inflow and outflow are higher than in other countries (see Table S2), mid-range movements seem to provide a better alternative to contact rates and short-range movements to partially explain time trends of COVID-19 cases and deaths (see Fig.~S14 of the SI).

From our analysis, we thus conclude that NETE values computed using contact rates as source time series are less sensitive to the province's geographic or demographic features, rather than to the noise of the target time series. 
Given good coverage, and consistency over time, contact rates thus represent a better epidemiological predictor of future COVID-19 deaths than short-range movements.

\section{Discussion}
In this work, we have introduced a novel framework based on transfer entropy to quantify the amount of information that is transferred from mobile phone-derived mobility metrics to epidemiological time series. 
Given the important role that mobility indicators have played in the COVID-19 pandemic, we tested our approach on mobility and epidemic time series collected in four European countries, between 2020 and 2021, at a subnational scale.
We found that, in general, the relative explanation added by mobility time series to predict future epidemic trends, whether new cases or deaths, was relatively small, ranging between 4\% and 6\% on average, and not statistically significant in the large majority of the provinces we considered, for any mobility metric. 
As a comparison, these values were about half of the relative explanation added by past knowledge of COVID-19 incidence to predict future deaths. 
Our method allowed us to directly compare the relative explanation added by different mobile phone-derived metrics of mobility: short- and mid-range mobility, and contact rates. 
We generally found a higher information transfer from contact rates than movement, in line with previous studies~\cite{crawford2022impact}, however, we also observed significant heterogeneities within the same country and between countries. 
With a classification model, we identified spatial features that may explain such heterogeneities. In provinces characterized by large populations, good coverage consistency over time, and small commuting in- and outflows, short-range movements can represent a useful metric to predict disease dynamics. 
Where commuting flows are large, such as in France, and Austria, mid-range movements, which represent less than 10\% of the total movements, provided a better alternative to short-range ones.  
Our results suggest the choice of the best mobility metric to inform epidemic predictions can depend on a number of different factors, even when using one single data provider. Moreover, our findings show that cell phone mobility metrics do not always capture epidemiologically-relevant behaviors and alternative data sources could be more effective for this aim, as, for instance, the collection of survey data~\cite{koher2022monitoring}.

There is an emerging common understanding that mobility indicators measured from mobile phone data present significant gaps and do not provide a consistent picture of mobility across countries, and data providers \cite{wardle2022gaps, gallotti2022distorting}. 
Previous studies have also highlighted the fact that coupling between mobility indicators and COVID-19 epidemiology is often weak, and it changes over time \cite{kishore2021evaluating}.
The approach we introduced here addresses the above challenges by providing a general framework to evaluate the quality of metrics derived from passively collected mobility traces as a predictor of epidemic outcomes. 
Our framework has the advantage of being model-free, meaning that it does not depend on modeling assumptions regarding the expected relationship between mobility and epidemic dynamics, nor it requires any parametrization. 
The normalized effective transfer entropy we adopted is a general method. 
It allows us to rigorously compare different mobility indicators, across epidemiological settings, by measuring the relative information added by mobility time series to the prediction of future disease incidence.
To this end, we release the code to reproduce our analysis between any two source and target time series (see Data and Code Availability).
Researchers can use this tool in any epidemiological context to gauge the added value of a specific mobile phone-derived behavioral measure for epidemic intelligence.   

Our study comes with a number of limitations and opens new directions for future work. 
We considered mobility metrics derived from one data provider, Meta, whose user base is not representative of the population in the countries we considered. 
However, alternative data sources of mobility indicators in Europe with a similar breadth, such as Google or Apple, do not reach the same spatial granularity and provide their data only as relative changes with respect to a pre-pandemic baseline, thus limiting their use in a study like ours. 
On the other hand, movement and colocation maps by Meta have been extensively used in several studies, including European countries \cite{bonaccorsi2020economic, galeazzi2021human, mazzoli2021projecting, smolyak2021effects, shepherd2021domestic}. 
Here, we considered four countries with different public health systems, and that adopted different testing strategies. 
Observed differences in the predictive power of mobility metrics across countries may depend on the varying quality of their reporting systems, especially at the province level. However, all four countries belong to the European Union and we expect  very similar standards of surveillance during the pandemic. 
Overall, it will be important to assess our findings on mobility data from other providers, and, most importantly, in countries of the non-Western world.
Finally, it is important to note that transfer entropy measurements become more accurate as the length of the source and target time series increases~\cite{marschinski2002analysing}. 
We worked with a relatively short time series, addressing the bias due to the small sample by adopting the effective transfer entropy. 
However, we could not systematically investigate how the information transfer changed over time, performing our analysis over different time windows and comparing them.
Future work could benefit from longer epidemic time series, over several years, to identify temporal changes in the information flow between human movements and COVID-19 dynamics. 

Measures of human mobility inferred from mobile phone data have been a critical ingredient to inform the public health response during the COVID-19 pandemic \cite{grantz2020use} and they will be an important asset in the fight against future pandemics. 
At the same time, their widespread use raises some relevant ethical concerns due to re-identification risks \cite{ienca2020responsible}, therefore, it is fundamental to assess the added value of using cell phone mobility data in a given epidemic scenario and whether the benefits outweigh the risks.
Our work provides a practical guide to identifying when and where mobile phone mobility metrics truly capture behavioral patterns that are relevant to predict disease dynamics.

\section{Materials and Methods}

\subsection{Epidemiological indicators}

We collected epidemiological time series in the 4 countries under study from 2 data sources. 
Daily reported cumulative COVID-19 cases were collected from the COVID-19 Data Hub \cite{guidotti2020covid}, an open source aggregator of up-to-date COVID-19 statistics, at the NUTS3 level in Austria, France, Italy, and Spain.

Daily reported cumulative deaths in Austria, France, and Spain were also collected from the COVID-19 Data Hub. 
For Italy, death statistics were only available on a weekly time scale from the public platform CovidStat (\url{https://covid19.infn.it/iss/}).

For the analysis, we generated daily incidence time series from cumulative data by computing day-to-day differences. 
Then, we further aggregated the daily time series of deaths and cases into weekly ones, to perform the transfer entropy analysis on a weekly scale.

\subsection{Mobility derived indicators}

In our study, we computed daily and weekly movement and contact rates from data provided by Meta through its Data for Good program \cite{Facebook_data}.
Here, we first describe the raw data sources provided by Meta and then the data processing we applied to compute the time series for the transfer entropy analysis.

\subsubsection{Raw data sources}
\label{sub:mob_data_sources}

We collected the following datasets that were publicly released by Meta since the beginning of the COVID-19 pandemic, in Austria, France, Italy, and Spain:
\begin{itemize}
    \item \textbf{Movement range maps}. It reports the number of users who moved between any two 16-level Bing tiles, with an 8-hour frequency. 
    \item \textbf{Users' population.} It reports the number of active users in each tile with an 8-hour frequency. The tile resolution is 4800 x 4800 $m^2$. 
    \item \textbf{Colocation maps.} It estimates the probability that, given any two administrative regions, $p_1$ and $p_2$, a randomly chosen user from $p_1$ and a randomly chosen user from $p_2$ are simultaneously located in the same place during a randomly chosen minute in a given week \cite{iyer2020large}. The dataset also reports the \textit{number of users} in $p_1$ and $p_2$. 
    \item \textbf{Stay put}. It reports for a given administrative region the daily percentage of users staying put within a single location, defined at the 16-level Bing tile.   
\end{itemize}
We formalize the description of the above datasets with the notation described in Table \ref{tab:data_original_time_series}: 

\begin{table}[h]
\begin{center}
\begin{adjustbox}{max width=\textwidth}
\begin{tabular}{llllll}
\toprule
{Dataset name} & $X_{s,t}$  & spatial resolution & temporal resolution & \\
\midrule
\textbf{population users} & $N_{t,h}^{(pop)}$            & t: tile (4800 x 4800 $m^2$) & h: 8 hour        &  \\
\textbf{movement between tiles} &  $M_{(t1,t2),h}$ & (t1,t2): tile pair (600 x 600 $m^2$) & h: 8 hour &  \\
\textbf{colocation probability} &   $P_{p,w}$ & p: province & w : week      &  \\
\textbf{colocation users} & $N_{p,w}^{(coloc)}$    & p: province & w: week      &  \\ 
\textbf{stay put} &     $S_{r,d}$  & r: region   & d: day       &  \\
\bottomrule
\end{tabular}
\end{adjustbox}
\caption{\label{tab:data_original_time_series} Summary of raw data sources as time series records $X_{s,t}$, where $s$ denotes the spatial resolution and $t$ the temporal resolution. 
}
\end{center}
\end{table}

\subsubsection{Aggregation of raw data}

\begin{table}[h]
\begin{center}
\begin{adjustbox}{max width=\textwidth}

\begin{tabular}{llllll}
\toprule
original data  & spatial aggregation & temporal aggregation & aggregated data & name\\
\midrule%
$N_{t,h}^{(pop)}$   & $\sum (t \in p)$   & $h$ interpolation and mean ($h \in w$) & $N_{p,w}^{(pop)}$  & \textbf{province population users} \\ 
$M_{(t1,t2),h}$     & $\sum$  $(t_1,t_2) \in p , t_1 = t_2$  & mean  ($h \in w$) & $M_{p,w}^{(within)}$  & \textbf{within tile province movement} \\
$M_{(t1,t2),h}$     & $\sum$ $(t_1,t_2) \in p , t_1 \neq t_2$    & mean ($h \in w$)  & $M_{p,w}^{(between)}$   & \textbf{between tiles province movement}  \\
$S_{r,d}$           &  $\forall p \in r \quad r=p$       & mean  ($d \in w$)  & $S_{p,w}$  & \textbf{province stay put} \\ 
\bottomrule
\end{tabular}
\end{adjustbox}
\caption{\label{tab:data_aggregated_time_series} Aggregation of data sources described in Table \ref{tab:data_original_time_series}, to generate our metrics of interest.}
\end{center}
\end{table} 

We then processed the raw data sources of Table \ref{tab:data_original_time_series} to obtain a set of time series having the same spatiotemporal resolution, that is weekly, at the NUTS3 scale. 
Results of the aggregation process are described in Table \ref{tab:data_aggregated_time_series}.
More in detail:
\begin{itemize}
    \item \textbf{Province users population.} (1) we performed a spatial aggregation by summing the population of tiles belonging to province $p$, thus obtaining a population at a (province, hour) level: $N_{p,h}^{(pop)}$. (2) we performed a linear interpolation of the temporal gaps that were present in $N_{p,h}^{(pop)}$ (3) we performed a temporal aggregation by averaging in each province, the 8h population records within a week.
    
    \item \textbf{Within tile province movement} (1) we first performed a temporal aggregation by averaging $M_{(t1,t2),h}$ for each pair $(t1,t2)$ over a week and obtaining $M_{(t1,t2),w}$ (2) we then performed a spatial feature joining and assigned each pair $(t1,t2)$ to the corresponding provinces $(p1,p2)$ (3) from $M_{(t1,t2),w}$ we obtained a \textbf{within tile province movement} $M_{p,w}^{(within)}$, that is the sum of movements which occurred in the same province $p$ and within the same tile.

    \item \textbf{Between tiles province movement} in the pipeline above, from step (3) we obtain a \textbf{between tile province movement}  $M_{p,w}^{(between)}$, that is the sum of movements which occurred in the same province $p$ and between two different tiles, $(t1,t2)$. By definition, the sum $M_{p,w}^{(between)} + M_{p,w}^{(within)}$ represents the total volume of movements in a province, in a week.  
    
    \item \textbf{Province stay put} (1) we performed a temporal aggregation on a weekly scale by performing the average and obtaining $S_{r,w}$ (2) we assign to each province $p$ the regional stay-put time series $S_{r,w}$ such that $p \in r$.   
\end{itemize}

\subsubsection{Computation of movement and contact rate}\label{mob_final_metrics}

We finally computed our metrics of interest, movement, and contact rates, as follows.
The short-range movement rate is defined as:
\begin{equation}
M_{p,w}^{s} = \frac{M_{p,w}^{(within)}}{N_{p,w}^{(pop)}}
\end{equation}
that is the proportion of users who moved within the same tile in a given province, in a given week. 
The mid-range movement rate is defined as:
\begin{equation}
M_{p,w} = \frac{M_{p,w}^{(between)}}{N_{p,w}^{(pop)}}
\end{equation}
representing the proportion of users who moved between different tiles in a given province, in a given week.
The contact rate is defined as:
\begin{equation}
CR(t)_{p,w} = \hat{P}_{p,w} \cdot N_{p,w}^{(pop)}
\end{equation}
where $\hat{P}$ denotes the colocation probability corrected by a factor that takes into account the overestimation of colocation probabilities due to the heterogeneous distribution of users across provinces and the presence of a significant fraction of static users in some periods of mobility restrictions \cite{mazzoli2021projecting} (see the SI for additional details). 

\subsubsection{Province sample selection}\label{sample}
The population of Facebook users who contribute to the generation of the movement and colocation time series varies across countries, and it changes over time. 
Moreover, the metrics of movement (short- and mid-range) and colocation, are computed from different users' samples of different sizes: $N_{p,w}^{(pop)}$ and $N_{p,w}^{(coloc)}$, respectively. 

In our analysis, to limit bias that may be caused by the little representativeness of the underlying sample of users, we selected NUTS3 regions in the 4 countries, according to the following criteria.
First, we considered only regions where the sample $N_{p,w}^{(pop)}$ represented at least $3\%$ of the census population to guarantee we had at least $500$ users in each province. 
Furthermore, we considered only those regions where the two sample sizes $N_{p,w}^{(pop)}$ and $N_{p,w}^{(coloc)}$ were always positively correlated over time, during the whole study period. 
We denote the Pearson's correlation of weekly values of $N_{p,w}^{(pop)}$ and $N_{p,w}^{(coloc)}$ as \emph{coverage consistency}. 

After the selection, our analysis includes 47 provinces in Austria, 51 provinces in France, 93 provinces in Italy, and 42 provinces in Spain, for a total of 233 spatial units.

Given two discrete temporal signals represented as time series $X$ and $Y$ the Transfer Entropy (TE) \cite{schreiber2000measuring} is a measure of the amount of information delivered from $X$ to $Y$, defined as: 
\begin{equation}
\label{eq_TE}
    TE_{XY} = H(Y|Y^{(l)}) - H(Y|Y^{(l)},X^{(l)}) \,,
\end{equation}
where $X^{(l)}, Y^{(l)}$ are respectively the $l$-lagged time series of $X$ and $Y$ and $TE_{XY}$ is formulated as a difference between two conditional entropy terms, where conditional entropy is expressed as $H(a|b) = H(a,b) - H(b)$, and $H(\cdot)$ is the Shannon Entropy. 
Given a discrete time series $S$, its observations can be expressed as the sample $\{ {s_i ; i=1,..,n} \}$, and we obtain the discrete probability distribution $p({s}_j)$. 
We compute the Shannon Entropy as: $H(S) = \sum_{j} p(s_j) \cdot log_2( p(s_j))$.
Thus $TE_{XY}$ can be expressed as:
\begin{equation}\label{TE_entropies}
    TE_{XY} = H(Y, Y^{(l)}) - H(Y^{(l)}) -  H(Y,Y^{(l)},X^{(l)}) + H(Y^{(l)},X^{(l)}).
\end{equation}
The time series that we consider in our experiments are continuous, therefore they need to be discretized before computing $TE_{XY}$. 
We employ the Kernel Density Estimation (KDE) for Transfer Entropy estimation. 
KDE method evaluates the entropy terms of Eq.\ref{TE_entropies} from the discretized density estimated from each of the four features sets: $\{(Y, Y^{(l)}), Y^{(l)}, (Y, Y^{(l)}, X^{(l)}), (Y^{(l)}, X^{(l)})\}$. 
KDE employs a Gaussian kernel for density estimation. Performing tests on synthetic datasets of different sizes, we checked this was the method the most adapted to small samples.
For the selection of the kernel's bandwidth, we use the Scott method \cite{scott2015multivariate}. 
The continuous density is then discretized with a grid obtained by an equal-width discretization of each feature's density domain. We select 20 as the number of bins for each feature's domain discretization. 
The discretized density is computed with the integral of the continuous probability density functions over each grid cell. 
Concerning the implementation, for TE estimation we use the PyCausality Python package (\href{https://github.com/ZacKeskin/PyCausality}{https://github.com/ZacKeskin/PyCausality}).

\paragraph{Effective Transfer Entropy.} 
We introduce the Effective Transfer Entropy (ETE) as a correction to TE for small sample time series, as originally proposed by \cite{marschinski2002analysing}: 
\begin{equation}\label{eq_ETE}
    ETE_{XY} = TE_{XY} - \frac{1}{N_{s}}\sum_{j=1}^{N_{s}} TE_{X\hat{Y_j}} \,,
\end{equation}
where the correction term is obtained by performing $N_{s}$ iterations of $Y$ shuffling, obtaining $\hat{Y_j}$ and computing the average of $\{ TE_{X\hat{Y_j}} ; j =1, ..,N_{s} \}$. In our experiments, we performed 500 shuffling iterations.

\paragraph{Normalized Transfer Entropy.} We would like to employ TE in order to compare a set of input signals $\{ X_j ; j=1,..,N\}$ in terms of their Transfer Entropy $TE_{X_jY}$ towards a specific output $Y$. From equation \ref{eq_TE} we have that $TE_{X_jY}$ is evaluated as a difference of conditional entropy where the first term $H(Y|Y^{(l)})$ depends only on target $Y$. 
In order to ensure comparability over the set $\{ TE_{X_jY} ; j=1,..,N\}$, we reformulate the difference as a relative difference dividing by $H(Y|Y^{(l)})$. Thus the set of inputs are compared according to $\{ TE_{X_jY}/H(Y|Y^{(l)}) ; j=1,..,N\}$ and we refer to $TE_{XY}/H(Y|Y^{(l)})$ as Normalized Transfer Entropy (NTE).  

\paragraph{Normalized Effective Transfer Entropy.} By combining the ETE and the NTE we can finally introduce the Normalized Effective Transfer Entropy (NETE), which is obtained by dividing the ETE by the first conditional entropy term $H(Y|Y^{(l)})$ as in \cite{perilla2012towards}: 
\begin{equation}\label{eq_NETE}
    NETE_{XY} = \frac{ TE_{XY} - \frac{1}{N_{s}}\sum_{j=1}^{N_{s}} TE_{X\hat{Y_j}} }{H(Y|Y^{(l)})}
\end{equation}
In this way, the NETE accounts both for bias in small sample time series and it ensures comparability between different input sources $\{X_j\}$ in terms of information transfer to different targets. Besides, it enables estimating the percentage of explanation value added with respect to only knowing the past of the time series used as a target.

\subsection{Classification model}
The introduction of the ETE allows associating a p-value, a metric of statistical significance, to each NETE value computed between any pair of time series.

In our study, we investigated a number of explanatory features to better understand why in some provinces the NETE could not identify a significant transfer of information between mobility time series and epidemiological indicators. 
More specifically, we trained a Random Forest classification model to predict the significance of $N_{X \rightarrow Y}$ at the threshold of $p<0.01$, in each province under study.
The random forest was performed with 100 decision tree classifiers on various sub-samples of the dataset and used averaging to improve the predictive accuracy and control for over-fitting. 
The function to measure the quality of a split was the Gini impurity.
Before applying the random forest, the data were split between training and test sets ($30\%$). 
To compensate for the imbalance of the datasets, we applied a Synthetic Minority Oversampling Technique \cite{chawla2002smote} on the test set.

As input to the classification model we used a set of features that characterize each province:
\begin{enumerate}
    \item population size (as reported by the latest available census);
    \item area (in km\textsuperscript{2});
    \item density of Facebook users (measured as $N_{p,w}$ divided by area);
    \item total cumulative number of reported COVID-19 deaths during the study period;
    \item commuting flow;
    \item coverage consistency; 
\end{enumerate}
The commuting flow is defined as the ratio between the total number of daily commuters who travel from or to a province and the total number of commuters who work and live in that province. Commuting data were collected from the latest available census statistics in each country.
The coverage consistency is the correlation over time between the users' populations $N_{p,w}^{(pop)}$ and $N_{p,w}^{(coloc)}$.

To quantify the importance of different features in our classification model, we used their SHAP (SHapley Additive exPlanations) values \cite{lundberg2017unified}. 
SHAP is a method to explain model predictions based on Shapley Values from game theory. In particular, we use TreeSHAP \cite{lundberg2018consistent}, an algorithm to compute SHAP values for tree ensemble models, such as the random forest classifier of our study.

\section{Data and code availability}
The data and code to reproduce our analysis are available at: \url{https://zenodo.org/record/7464949#.Y6L0CfxKhNg}

\section{Funding}
F.D. gratefully acknowledges support from the CRT Lagrange Fellowships in Data Science for Social Impact of the ISI Foundation, where this work was conducted.
M.T. and L.G. acknowledge the Lagrange Project of the ISI Foundation funded by CRT Foundation.
The funders had no role in the study design, decision to publish, or preparation of the manuscript.

\section{Acknowledgements}
We gratefully acknowledge Alex Pompe for his help to understand the details of mobility data from Meta.

\section{Author contributions}
FD collected data, conducted experiments, interpreted the results, made figures, and contributed to the writing of the paper.
MT and LG conceived and designed the study, conducted the statistical analysis, interpreted the results, made figures, and wrote the paper. 
All authors read and approved the final version of the manuscript.

\section{Competing interests}
The authors declare no competing interests.

\setcounter{page}{1}
\renewcommand{\thetable}{S\arabic{table}}
\setcounter{figure}{0}
\renewcommand{\thefigure}{S\arabic{figure}}

\newpage

\paragraph{Supplementary Information}
\section*{Correction to the colocation probability}

Colocation maps provided by Meta is defined as the number of colocation events over the number of possible events. This, by design, includes interactions between users staying within the same tile but not having actual contact with other users. 
For this reason, we estimate the contact rate in each province by removing the contribution due to the users staying put. 
We explain our approach to estimating such contribution in the following. 
Let us start by writing the original colocation probability $P$ as: 
\begin{equation}\label{dim_pc_0}
    P = \frac{E}{N^2}
\end{equation}
where: 
\begin{itemize}
    \item $E$ is the number of colocation events within the province
    \item $N$ is the number of province colocation users.
\end{itemize}
The exact formula should be $P = \frac{E}{N(N-1)}$ but as $N$ is large we approximate it to \ref{dim_pc_0}.
Let us denote $R^{(c)}$ the number of measured colocation events that are due to users who stay put only, then the corrected colocation probability should be written in the following way:
\begin{equation}
    \hat{P}_{p,w} = \frac{E - R^{(c)}}{ N^2 }
\end{equation}
We estimate $R^{(c)}$ by using the stay-put probability $S$, which is the probability of a user staying put. 
Let us call the tile population ratio probability distribution $\{ f_{t_l} ; t = 1,.., T_l \}$ where $T$ is the number of tiles in a province.
This gives us an estimate of the contribution of the users who stay put to the colocation probability, as:
\begin{equation}
  R^{(c)}= \sum_{t=1}^{T} N^2\cdot f_t^2 \cdot S^2 .
\end{equation}
So we rewrite: 
\begin{equation}\label{prob_coloc_v2}
   \hat{P}_{p,w} = P - S^2 \sum_{t=1}^{T_l} f_{t_l}^2 
\end{equation}
We do not have access to the population of the tiles used for the colocation so we make an approximation using the population distribution given for each tile with dimensions $4800$ m $\times$ 4800 m.
As there are by definition 64 colocation tiles within a single population tile, the expression  Eq.\ref{prob_coloc_v2} can be formulated as: 
\begin{equation}\label{prob_coloc_v3_first}
     \hat{P}_{p,w} = P_{p,w} - S_{p,w}^2 \cdot \sum_{t=1}^{T} 64 \cdot \left( \frac{f_{t,w}^{(p)}}{64} \right)^2 
\end{equation}
where:
\begin{itemize}
  \item $f_{t,w}^{(p)} = \frac{N_{t,w}}{N_{p,w}} ; t \in p$ : tile $t$ population frequency in province $p$.
    \item $N_{t,w}$ : population at (tile,week) resolution. It is obtained through mean temporal aggregation of $N_{t,h}$ over the week interval denoted by $w$.
    \item $N_{p,w}$ : population at (province, week) resolution. It is obtained through sum spatial aggregation of $N_{t,w}$ over the tiles belonging to province $p$.
    \item $T$ is the number of tiles $4800$ m $\times$ 4800 m
\end{itemize}
We can introduce the quantity $Q_{p,w}$ as the sum of squared frequencies of the province tile distribution $Q_{p,w} = \sum_{t \in p} (f_{t,w}^{(p)})^2 $, so that, finally:
\begin{equation}\label{prob_coloc_v3_second}
     \hat{P}_{p,w} = P_{p,w} - \frac{S_{p,w}^2 \cdot Q_{p,w}}{64}
\end{equation}

\begin{table}[tb]
\centering
\begin{tabular}{lcc}
\hline
 &  $M^s$ (\%) &  $M$ (\%) \\
\hline
Austria &  99.6 [97.9 -- 100]  & 0.5 [0.0 - 2.1] \\
France & 91.3 [88.2 -- 93.7]   & 8.8 [6.3 -- 11.9]   \\
Italy & 89.9 [86.4 -- 92.8] &  10.2 [7.2 -- 13.6] \\
Spain & 91.5 [86.1 -- 95.1] & 8.5 [4.9 -- 13.9]  \\
\hline
\end{tabular}
\label{tab_SI:mobility}
\caption{\textbf{Relative proportion of mobility components in each country.} Each row displays the proportion of movements, as a percentage of the total movements within each province, that are represented by the short-range mobility ($M^s(t)$) and the mid-range mobility ($M(t)$). Each table entry reports the median value and the IQR, computed over all provinces, and all weeks of the study period. Short-range mobility represents the large majority of movements within a province, in all countries.}
\end{table}

\begin{table}[tb]
\centering
\begin{tabular}{lcc}
\hline
{} & coverage consistency & commuting flow \\                  
\hline
Austria &  0.64 [0.45--0.79] &  1.05 [0.43--1.69] \\
France  &  0.32 [0.23--0.43] &  0.30 [0.22--0.51] \\
Italy   &  0.63 [0.42--0.77] &  0.21 [0.12--0.29] \\
Spain   &  0.86 [0.68--0.91] &  0.08 [0.05--0.10] \\
\hline
\end{tabular}
\label{table_char}
\caption{\textbf{Coverage consistency and commuting flow distributions by country.} Each table entry reports the median value and the IQR  computed over all provinces, in each country, considered in the study.}
\end{table}

\begin{figure}[h]
\centering
\includegraphics[width=0.8\textwidth]{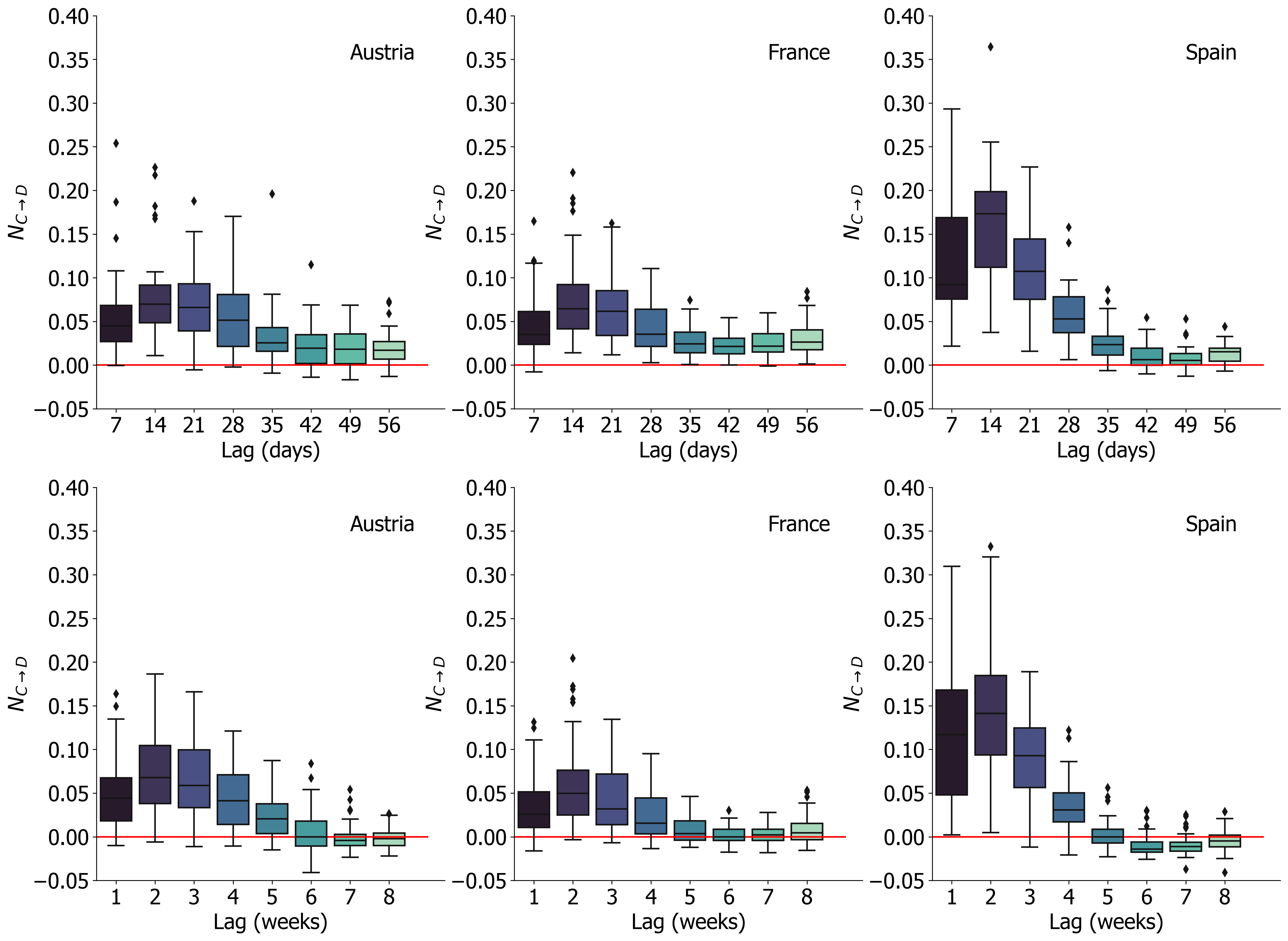}
\caption{\label{fig:daily} \textbf{Comparison of NETE values computed on weekly and daily time series.} $N_{C \rightarrow D}$ computed between time series data collected on a weekly time scale (bottom row) and a daily one (top row).
Daily time series were available only for Austria, France and Spain.}
\end{figure}


\begin{figure}[h]
\centering
\includegraphics[width=0.8\textwidth]{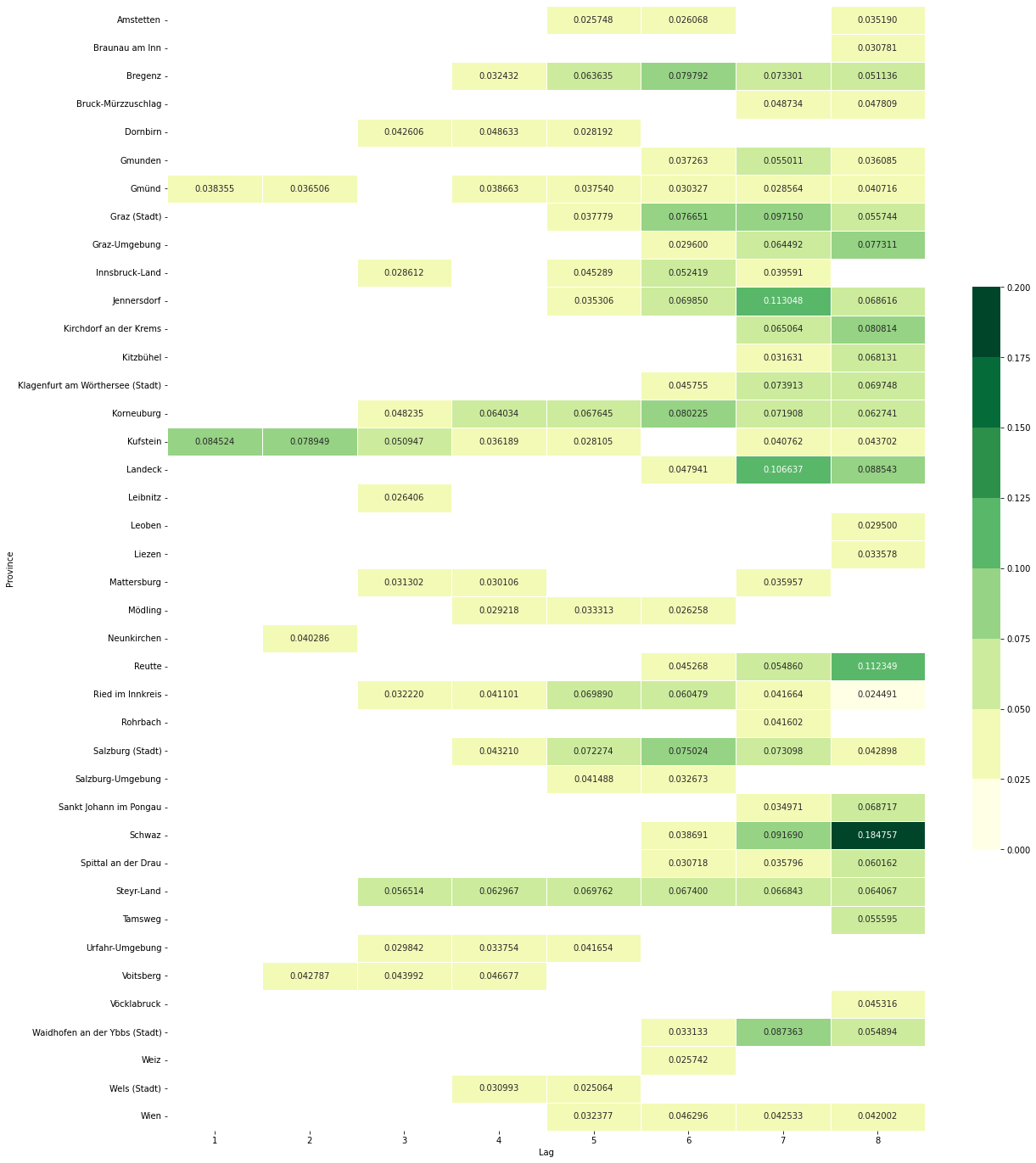}
\caption{\label{fig:CRA} \textbf{NETE values from contact rates to deaths in Austria.} Only statistically significant values are shown (p-value$<0.01$).}
\end{figure}

\begin{figure}[h]
\centering
\includegraphics[width=0.8\textwidth]{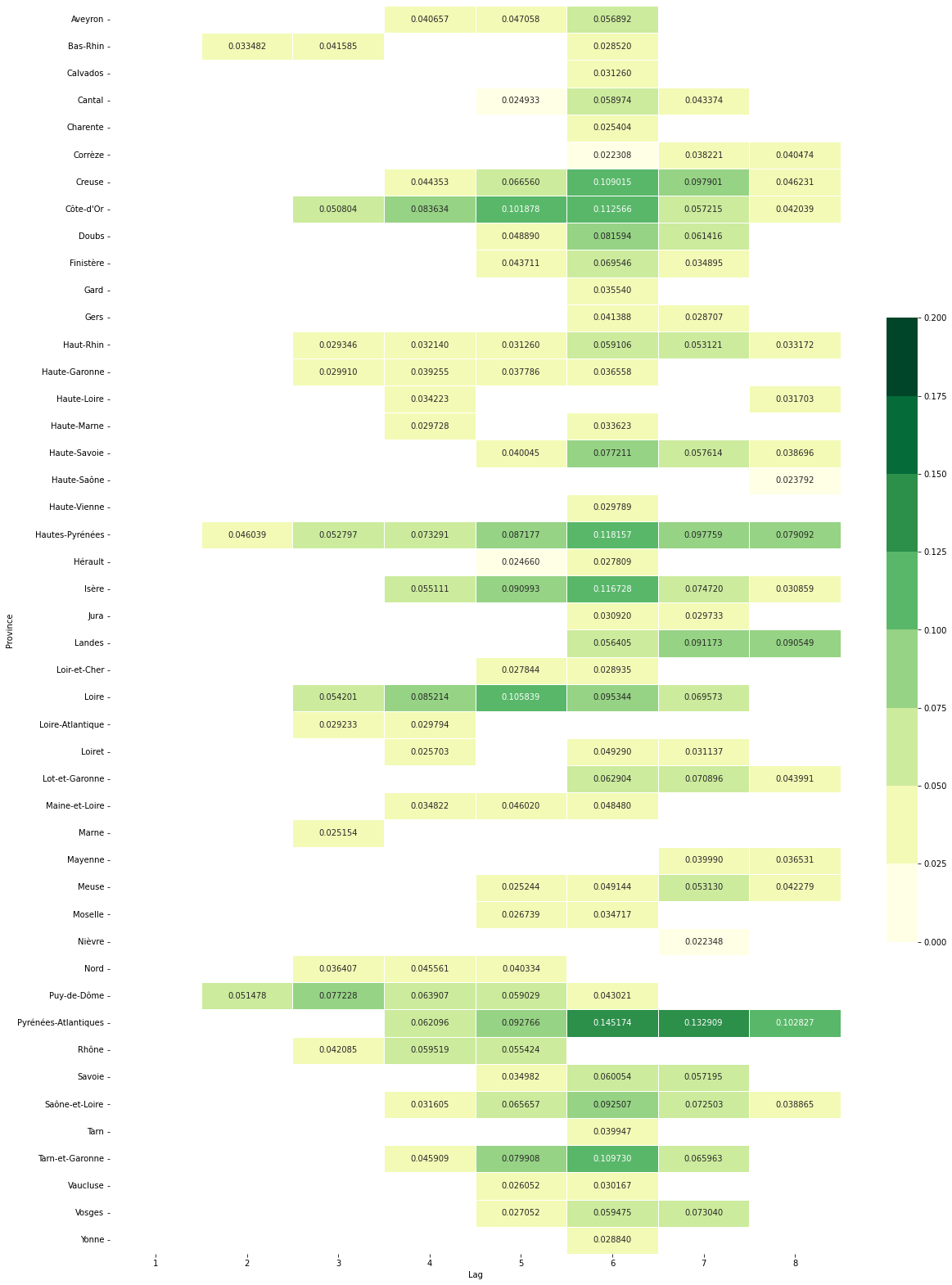}
\caption{\label{fig:CRF} \textbf{NETE values from contact rates to deaths in France.} Only statistically significant values are shown (p-value>0.01).}
\end{figure}

\begin{figure}[h]
\centering
\includegraphics[width=0.8\textwidth]{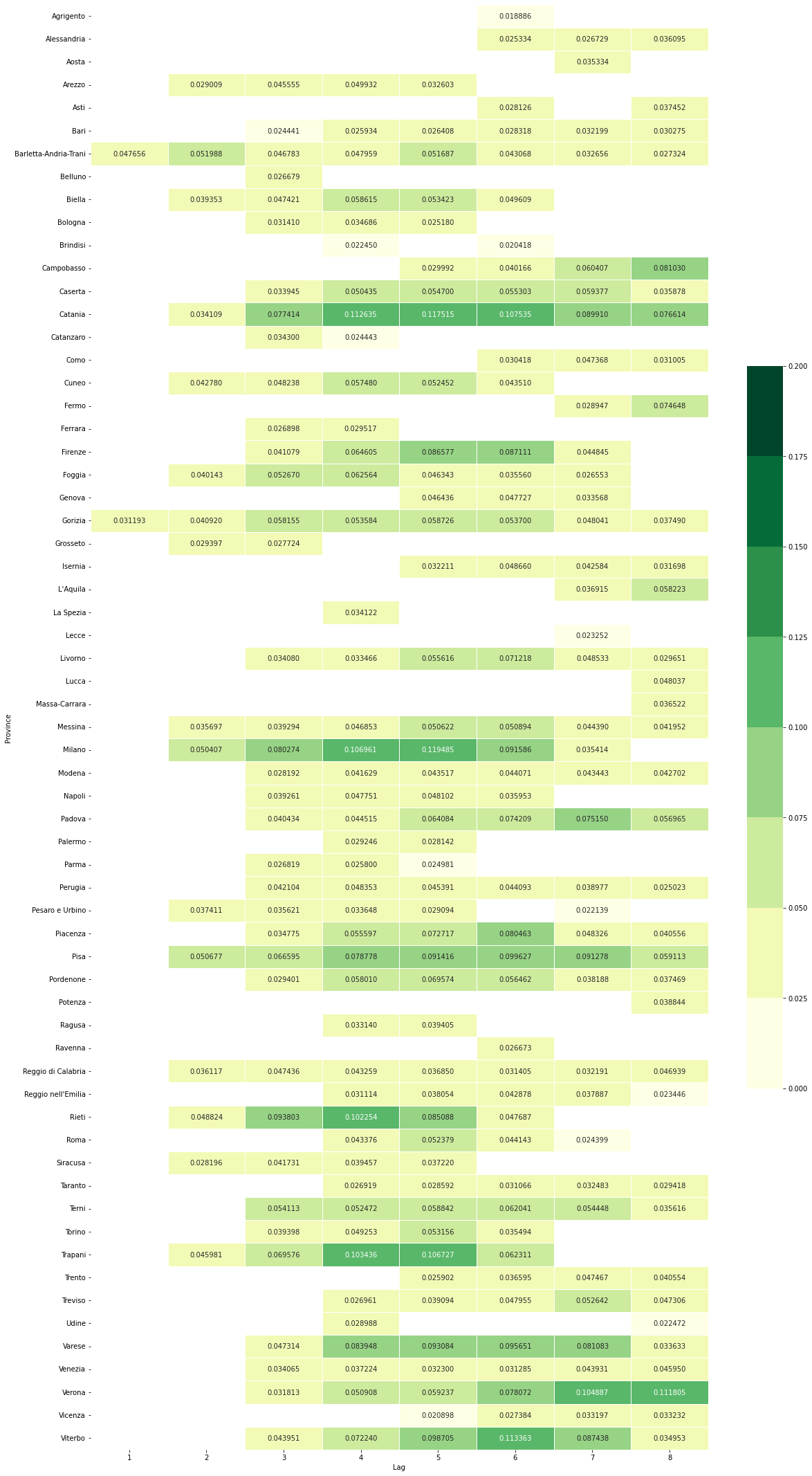}
\caption{\label{fig:CRI} \textbf{NETE values from contact rates to deaths in Italy.} Only statistically significant values are shown (p-value$<0.01$).}
\end{figure}

\begin{figure}[h]
\centering
\includegraphics[width=0.8\textwidth]{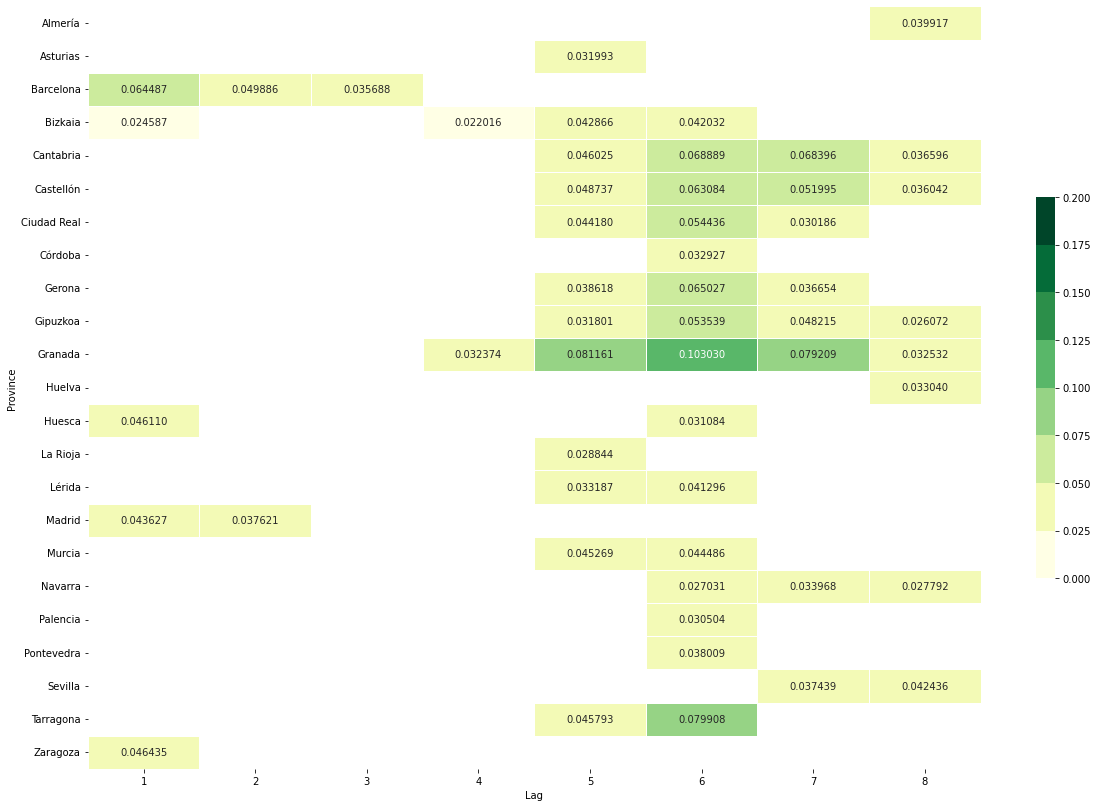}
\caption{\label{fig:CRS} \textbf{NETE values from contact rates to deaths in Spain.} Only statistically significant values are shown (p-value$<0.01$).}
\end{figure}

\begin{figure}[h]
\centering
\includegraphics[width=0.8\textwidth]{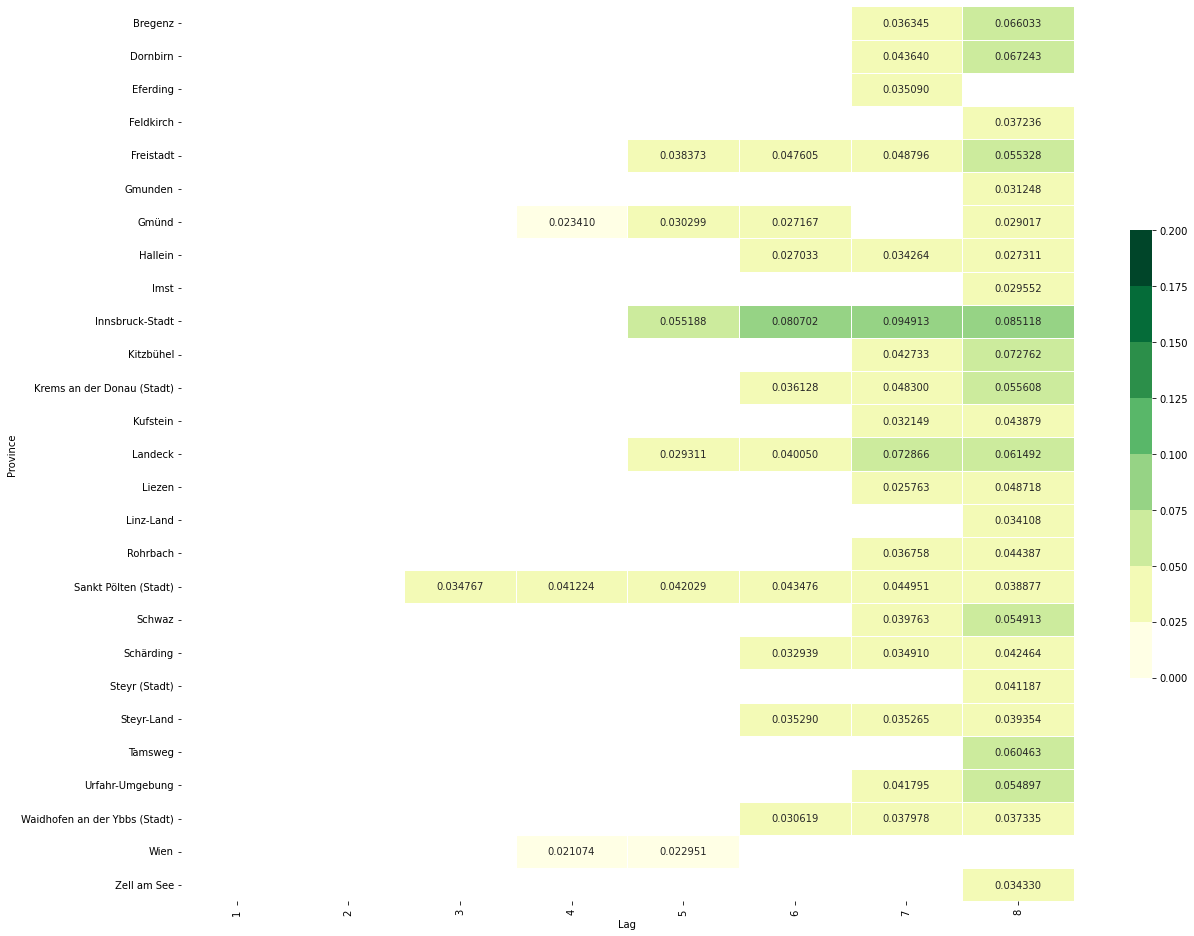}
\caption{\label{fig:MVA} \textbf{NETE values from movements to deaths in Austria.} Only statistically significant values are shown (p-value$<0.01$).}
\end{figure}

\begin{figure}[h]
\centering
\includegraphics[width=0.8\textwidth]{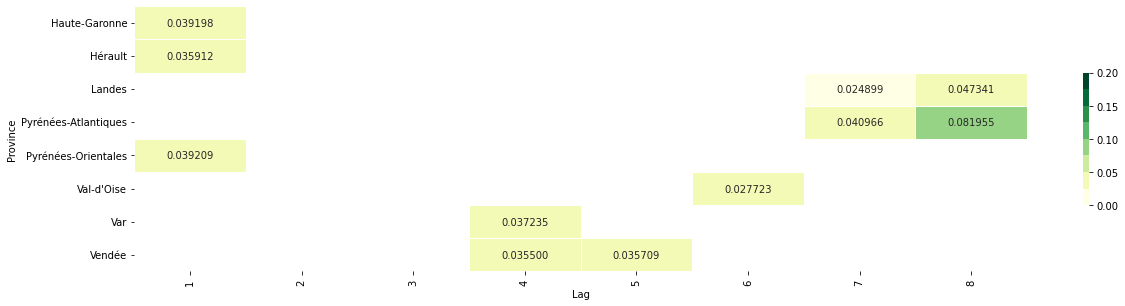}
\caption{\label{fig:MVF} \textbf{NETE values from movements to deaths in France.} Only statistically significant values are shown (p-value$<0.01$).}
\end{figure}

\begin{figure}[h]
\centering
\includegraphics[width=0.8\textwidth]{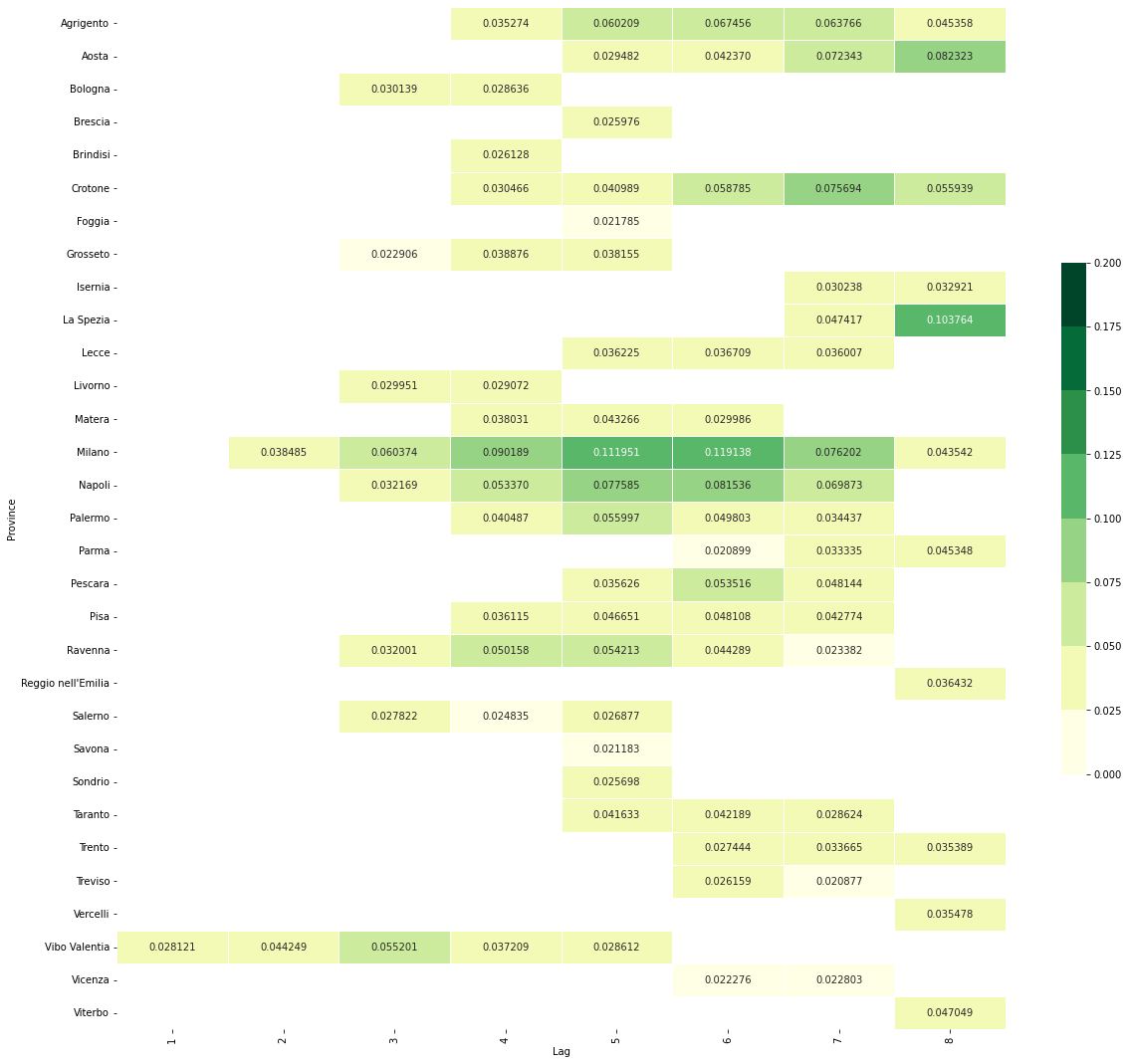}
\caption{\label{fig:MVI} \textbf{NETE values from movements to deaths in Italy.} Only statistically significant values are shown (p-value$<0.01$).}
\end{figure}

\begin{figure}[h]
\centering
\includegraphics[width=0.8\textwidth]{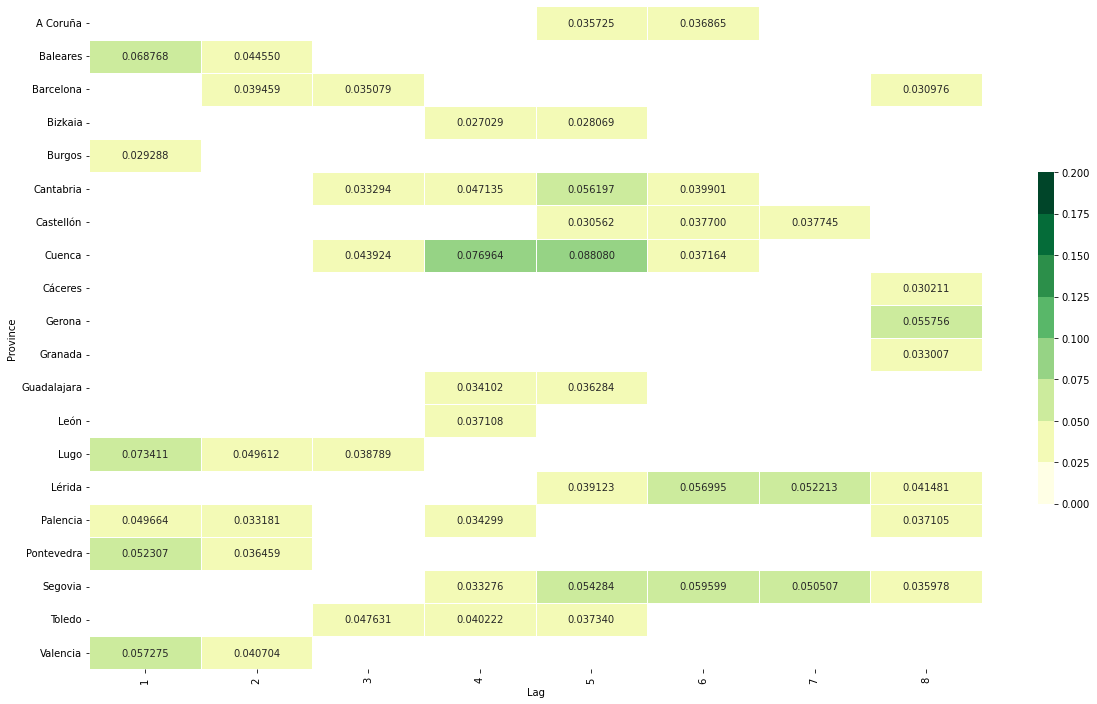}
\caption{\label{fig:MVS} \textbf{NETE values from movements to deaths in Spain.} Only statistically significant values are shown (p-value$<0.01$).}
\end{figure}
\begin{figure}[h]
\centering
\includegraphics[width=0.8\textwidth]{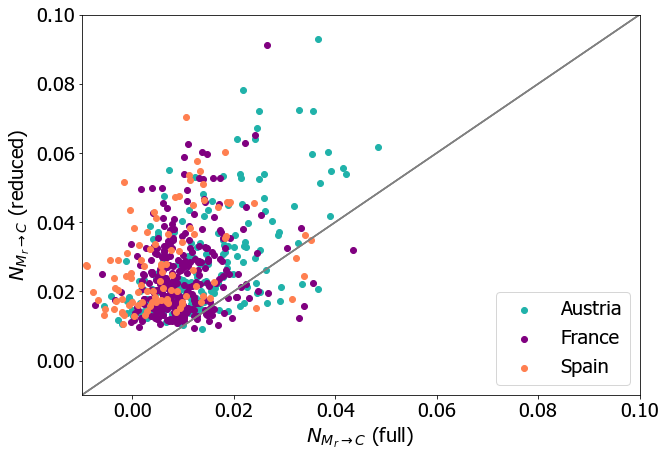}
\caption{\label{fig:trim_vax} \textbf{Comparison of NETE values computed on full time series and reduced time series.} $N_{M \rightarrow C}$ computed between time series data collected including the vaccination campaign (full) and not (reduced).
The reduced study period ranges from September 1, 2020 to January 31, 2021. The full study period extends up to July 31, 2021.
We consider daily time series only to address biases due to small samples.}
\end{figure}

\begin{figure}[tb]
\centering
\includegraphics[width=\textwidth]{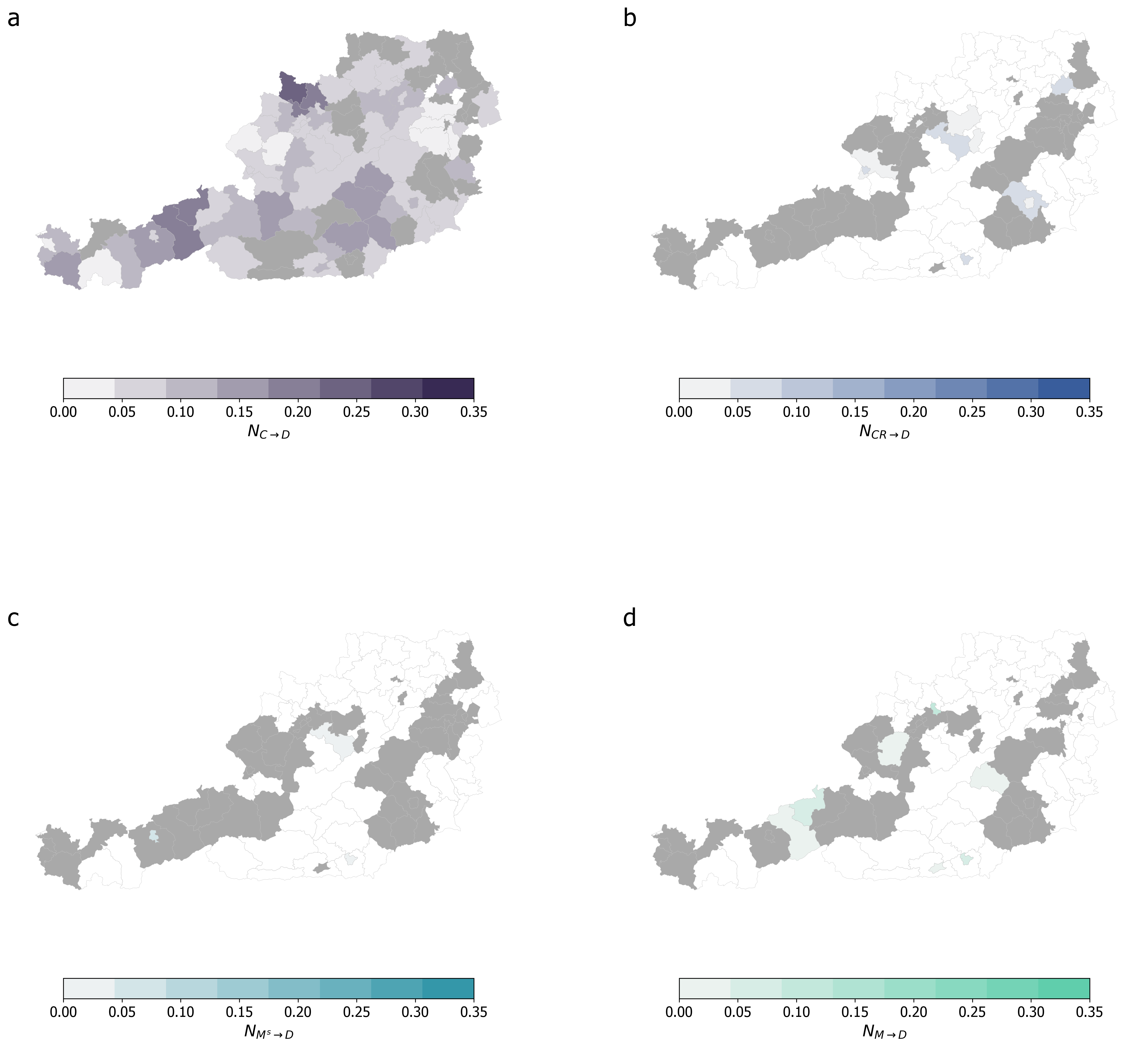}
\caption{\label{fig:map_Austria} \textbf{Spatial variations of normalized effective transfer entropy.} Maps of NETE values computed for different source time series and weekly COVID-19 deaths, in the provinces of Austria: (a) source is COVID-19 cases at lag $l$=2 weeks, (b) source is contact rate at lag $l$=7 weeks, (c) source is short-range movement at lag $l$=7 weeks. (d) source is mid-range movement at lag $l$=7 weeks.
Dark grey indicates provinces with non-significant values of NETE ($p>0.01$). Provinces in white are excluded from our sample. 
}
\end{figure}

\begin{figure}[tb]
\centering
\includegraphics[width=\textwidth]{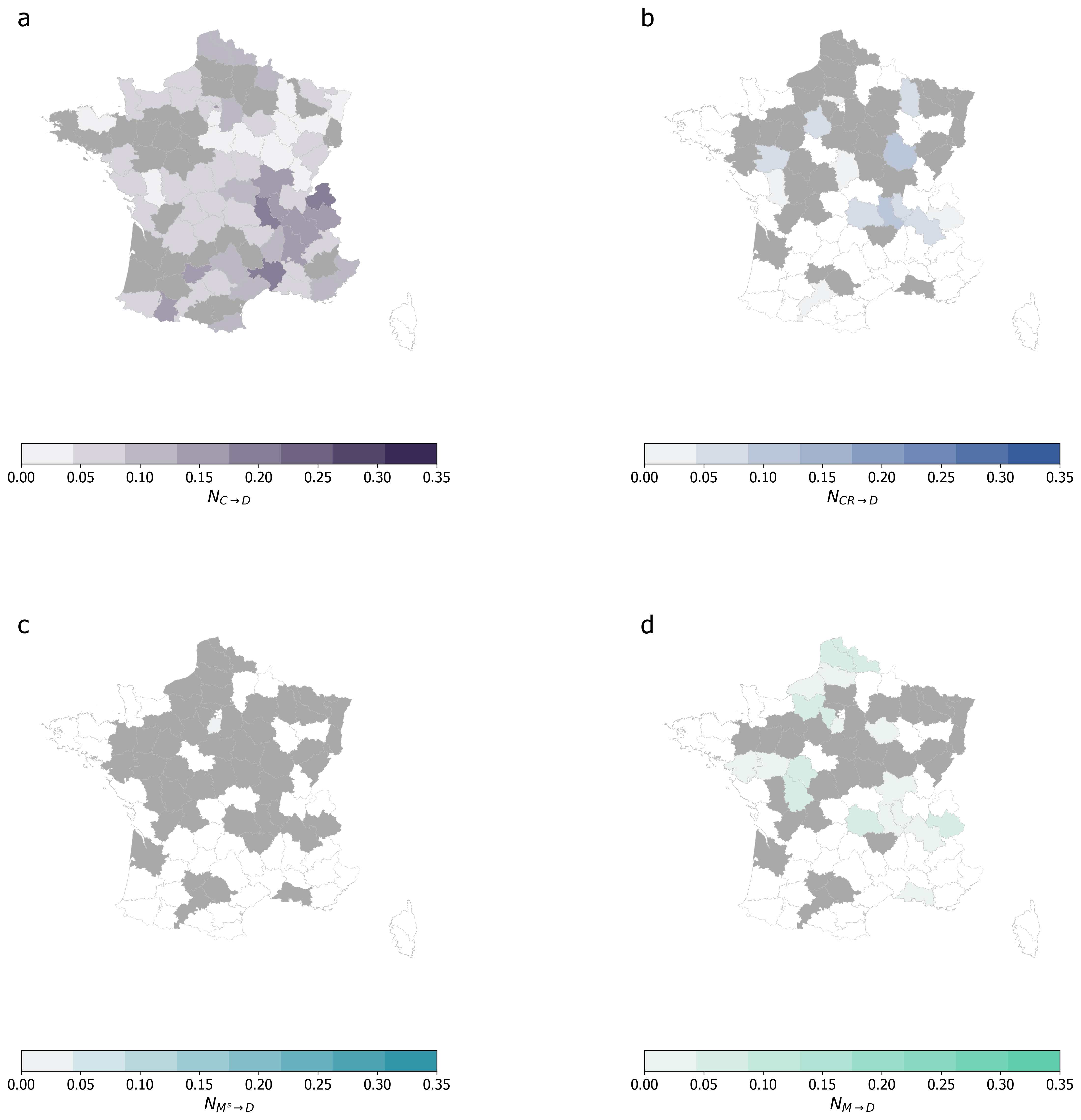}
\caption{\label{fig:map_Austria} \textbf{Spatial variations of normalized effective transfer entropy.} Maps of NETE values computed for different source time series and weekly COVID-19 deaths, in the provinces of France: (a) source is COVID-19 cases at lag $l$=2 weeks, (b) source is contact rate at lag $l$=7 weeks, (c) source is short-range movement at lag $l$=7 weeks. (d) source is mid-range movement at lag $l$=7 weeks.
Dark grey indicates provinces with non-significant values of NETE ($p>0.01$). Provinces in white are excluded from our sample. 
}
\end{figure}

\begin{figure}[tb]
\centering
\includegraphics[width=\textwidth]{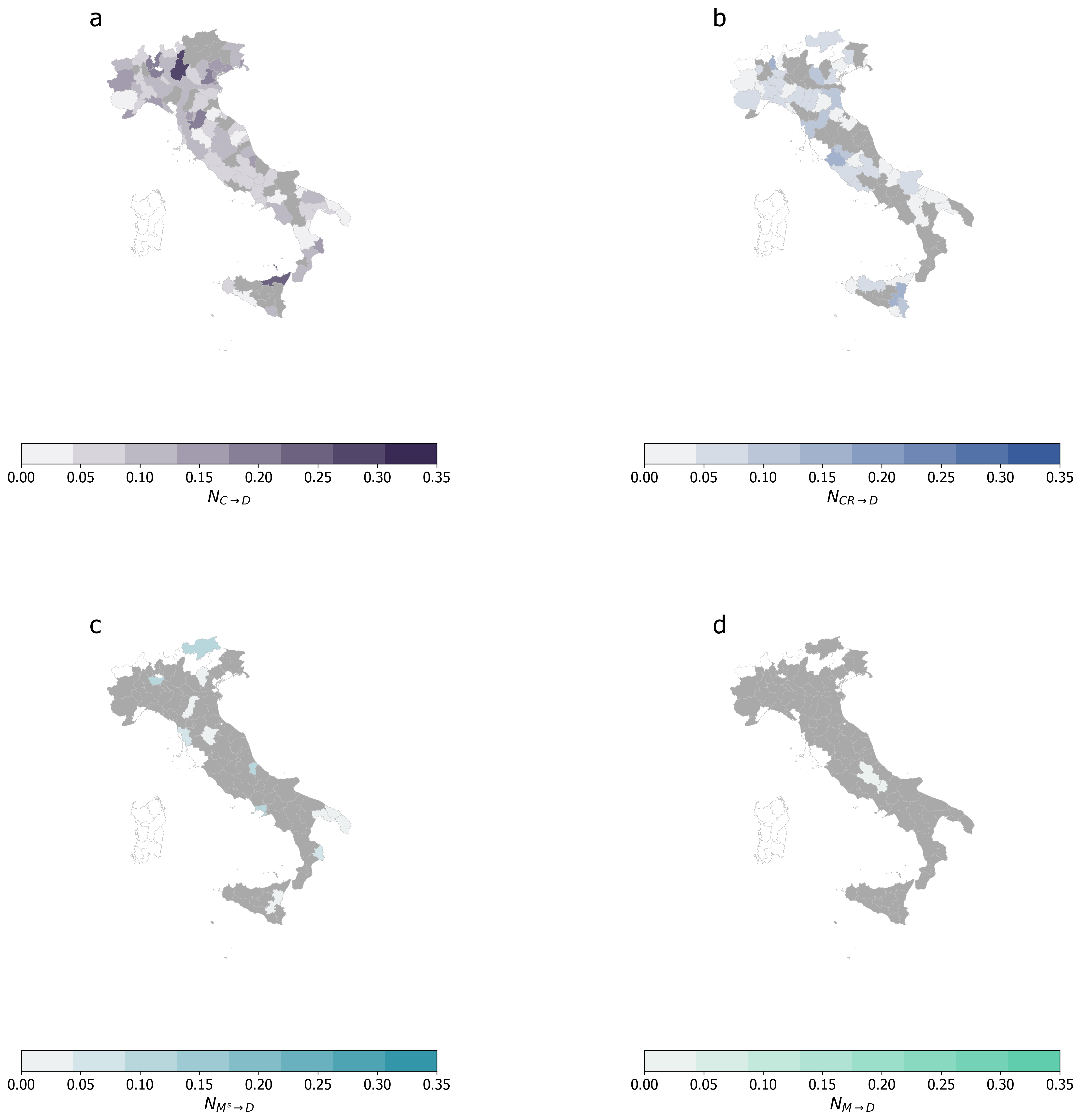}
\caption{\label{fig:map_Italy} \textbf{Spatial variations of normalized effective transfer entropy.} Maps of NETE values computed for different source time series and weekly COVID-19 deaths, in the provinces of Italy: (a) source is COVID-19 cases at lag $l$=2 weeks, (b) source is contact rate at lag $l$=7 weeks, (c) source is short-range movement at lag $l$=7 weeks. (d) source is mid-range movement at lag $l$=7 weeks.
Dark grey indicates provinces with non-significant values of NETE ($p>0.01$). Provinces in white are excluded from our sample. 
}
\end{figure}

\begin{figure}[tb!]%
    \centering
    \includegraphics[width=\textwidth]{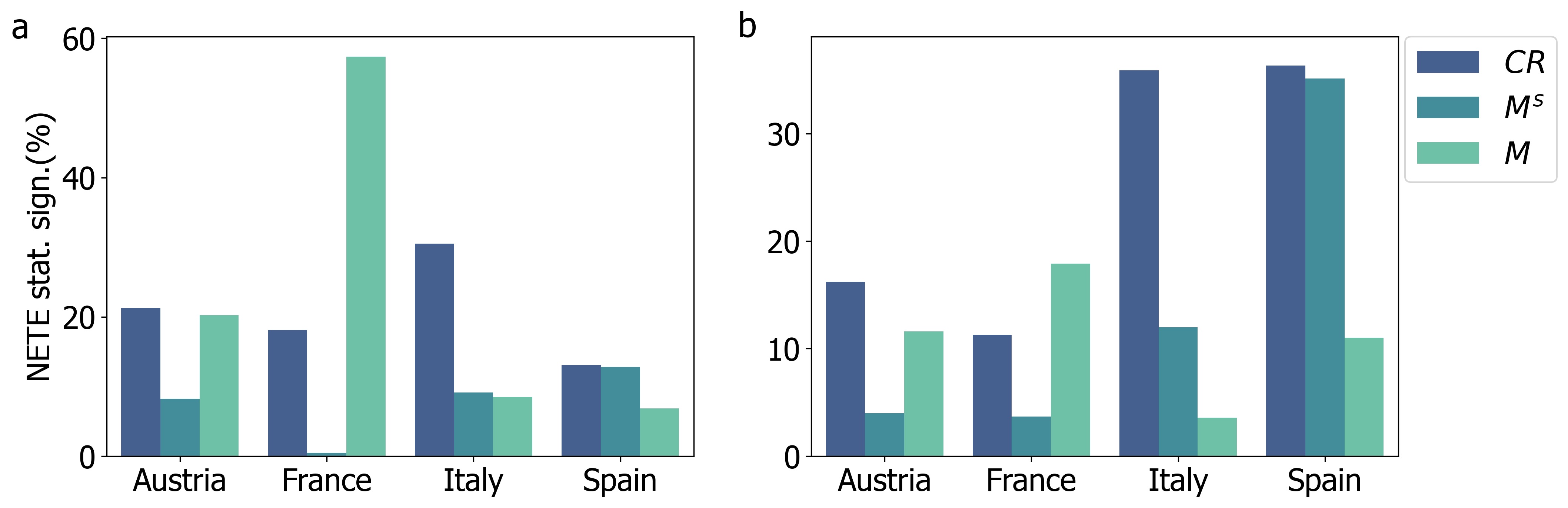}%
    \caption{\textbf{Percentage of statistically significant NETE values, disaggregated by country and by mobility metric used as source variable.} Target variables are: weekly COVID-19 cases (panel a) and weekly COVID-19 deaths (panel b).}%
    \label{fig:metrics_country}%
\end{figure}
\bibliographystyle{unsrt}


\end{document}